\documentclass{aa} 
\usepackage{lineno}
\usepackage{url}
\usepackage{amssymb}
\usepackage{epstopdf}
\usepackage{color}
\usepackage[usenames,dvipsnames,svgnames,table]{xcolor}
\usepackage{graphicx}
\usepackage{caption}
\usepackage{subcaption}
\usepackage{txfonts}
\usepackage{natbib}
\begin{document} 

\titlerunning{On the accretion process in a high-mass star forming region}

\title{On the accretion process in a high-mass star forming region}

\subtitle{A multitransitional THz \emph{Herschel}\thanks{Herschel is an ESA space observatory with science instruments provided by European-led Principal Investigator consortia and with important participation from NASA}-HIFI   study of  ammonia toward  G34.26+0.15}

\authorrunning{M.~Hajigholi et~al.}   
\author{
M.~Hajigholi\inst{1}, 
C.M.~Persson\inst{1}, 
E.S.~Wirstr\"om\inst{1}, 
J.H.~Black\inst{1}, 
P.~Bergman\inst{1},  
A.O.H.~Olofsson\inst{1},
M.~Olberg\inst{1}, 
F.~Wyrowski\inst{2},
A.~Coutens \inst{3},
\AA.~Hjalmarson\inst{1}, and
K.M.~Menten \inst{2}
}

   \institute{Chalmers University of Technology, Department of Earth and Space Sciences, Onsala Space Observatory,  SE-439\,92 Onsala, Sweden \\ \email{\url{mitra.hajigholi@chalmers.se}}  \and Max-Planck-Institute f\"ur Radioastronomie, Auf dem H\"ugel 69,
53121 Bonn, Germany  
\and Centre for Star and Planet Formation, Niels Bohr Institute \& Natural History Museum of Denmark, University of Copenhagen, {\O}ster Voldgade 5-7, DK-1350 Copenhagen K., Denmark
}

\date{Received 1 May, 2015 /  Accepted 13 October, 2015}
   
\abstract
{}
{Our aim is to explore the gas dynamics  and the accretion process in the early phase of high-mass star formation. }
{The inward motion of molecular gas in the massive star forming region G34.26+0.15 is investigated by using high-resolution profiles of seven transitions of ammonia at THz frequencies observed with \emph{Herschel}-HIFI.  The shapes and intensities of these lines are interpreted in terms of radiative transfer models of a spherical, collapsing molecular envelope. An accelerated Lambda Iteration (ALI) method is used to compute the models.}
{The seven ammonia lines show mixed absorption and emission with inverse P-Cygni-type profiles that suggest infall onto the central source. 
A trend toward absorption at increasingly higher velocities for higher excitation transitions is clearly seen in the line profiles. 
The $J = 3\leftarrow2$ lines show only very weak emission, so these absorption profiles can be used directly  to analyze the inward motion of the gas. This is the first time a multitransitional study of spectrally resolved rotational ammonia lines has been used for this purpose. Broad emission is, in addition, mixed with the absorption in the  $1_0-0_0$ ortho-NH$_3$ line, possibly tracing a molecular outflow from the star forming region. 
The best-fitting ALI model reproduces the continuum fluxes and line profiles, but slightly underpredicts the emission and absorption depth in the ground-state ortho line $1_0-0_0$. 
An ammonia abundance on the order of 10$^{-9}$ relative to H$_2$ is needed to fit the profiles.
The derived ortho-to-para ratio is approximately 0.5 throughout the \textnormal{infalling} cloud core similar to recent  findings for translucent clouds in sight lines toward W31C and W49N.
We find evidence of \emph{two} gas components moving inwards toward the central region with constant velocities:  2.7 and  5.3~km$\,$s$^{-1}$, relative to the source systemic velocity.
Attempts to model the inward motion with a single gas cloud in free-fall collapse did not succeed. The inferred mass accretion rates derived rises from \mbox{$4.1\times10^{-3}$} to $\mbox{$4.5\times10^{-2}$~M$_{\odot}$~yr$^{-1}$}$, which is sufficient to overcome the expected radiation pressure from G34.26+0.15.
}
{} 
 

\maketitle

\section{Introduction}\label{introduction} 
This paper presents a detailed observational analysis of the early and rapid mass-accretion/gravitational-inflow phase necessary for forming a massive protostellar cluster, subsequently leading to the formation of high-mass stars. We therefore begin with an overview of the star formation scenario and related problems (for more details see \citealt{McKee2007}).

Massive stars ($\gtrsim8\, M_{\odot}$) are known to be born in hot ($\gtrsim100$~K), compact ($\lesssim0.1$~pc), and dense cores within giant molecular clouds. 
After a massive young stellar object (MYSO) forms in a collapsing core, its ultraviolet radiation dissociates molecular hydrogen (H$_{2}$) and ionizes the resulting atomic hydrogen to form an \ion{H}{II} region. 
In the earliest stages, the surrounding gas core is dense enough to slow down the ionization front, which leads to the formation of an ultra compact \ion{H}{II} (UC~\ion{H}{II}) region, a key phase in the early lives of massive protostars \citep{Hoare2005, Beuther2007, Zinnecker2007}. Many UC~\ion{H}{II} regions are found to be associated with warm molecular gas as proven by highly excited ammonia  \citep{cesaroni1992}, carbon monosulfide \citep{olmi1999},  methanol masers \citep{walsh1998}, and other molecular tracers of the early hot core phase \citep{hatchell1998}. 
The high column density of the gas and dust that surrounds the UC~\ion{H}{II} regions  makes them wholly invisible at UV and visible wavelengths, but their high luminosities ($10^4-10^6$ $L_\odot$) dominate the appearance of star forming regions at far-infrared wavelengths. The emission stems from heated dust within the ionized gas and the dense surrounding molecular envelope, and is identified by its steeply rising spectrum from near- to mid-infrared wavelengths.

The most challenging goal in massive star formation is to explain how a protostar can accumulate a large amount of infalling mass, despite its strong radiation pressure. 
MYSOs show bipolar molecular outflows that require ongoing accretion from infalling material with high angular momentum from the surrounding molecular core. 
Models considering a protostar-disk system \citep[e.g., ][]{York, Krumholz05, Banerjee, Tan2002, McKee2003, Bonnell2006, Padoan2014} suggest how the accretion of matter can overcome radiative pressure. 
Evidence of such accretion has been observed toward low-mass star formation cores \cite[e.g.,][]{Chou2014}, but similar observational evidence is still scarce toward high-mass star forming regions.
Not only are MYSOs typically observed at large distances, have high extinction, and form in clusters, but they also have relatively short time scales on their evolutionary phases because high far-UV and extreme UV fluxes photo evaporate the disk on time scales of $\sim10^5$~yrs. The disk can  be observable indirectly as a deeply embedded UC~\ion{H}{II} region with a comparable lifetime \citep{Zinnecker2007}.  

To trace the dynamics of gas in the deeply embedded phase of star formation, resolved molecular emission and absorption line profiles must be used. 
The high dust column densities toward massive star forming (MSF) regions result in strong continuum radiation,  which can be used as background sources for absorption line studies at THz frequencies. A redshifted absorption profile relative to the systemic velocity of the star forming core is a direct probe of inward gas motion and can be used through all embedded stages of massive star formation. This method is much more reliable than studies based on observations of the so-called blue-skewed \textnormal{emission} line profiles with a   central  self-absorption and a  blue peak stronger than the red peak. At kpc distances, these profiles might be mistaken for infall instead of other kinematics such as rotation and outflow activities.

In the earliest and coldest stages of molecular clumps, nitrogen-containing molecules such as ammonia can be used as tools for studies of massive star formation, as they are known  to survive in the gas phase when other species freeze out onto cold dust surfaces \citep{Bergin}.
The spectroscopic properties of ammonia (\element[][][][3]{NH}) have proven to be very useful as a probe of a wide range of physical conditions in a variety of interstellar environments \citep{ho}. One reason is ammonia's wide range of Einstein $A$-coefficients (and therefore of critical densities, $n_{\rm crit}$) covered by its many transitions (see Table~\ref{Table: transitions}), but also since it is a classical symmetric top configuration, which makes it sensitive to temperature. 

Ammonia inversion transitions at cm wavelengths have been widely observed toward bright UC~\ion{H}{II} regions, for example, by \citet{sollins2005}, \citet{Zhang}, and \citet{Beltran} in G10.6-0.4, W51, and G24.78+0.08. These observations indicated the presence of inward gas motion as predicted by the accretion model. This method, however,  traces a fairly late stage of massive star formation as it requires an already developed UC \ion{H}{II} region to be used as a background source at these wavelengths. Pure rotational transitions at THz frequencies can, on the other hand, be observed in absorption against the strong thermal emission from the heated dust in the direction of the MSF regions.  

Rotational  transitions of ammonia at THz frequencies are not possible to observe 
using ground-based observatories due to the opaque terrestrial atmosphere. Some observations of the ground state rotational transition of ortho-\element[][][][3]{NH}  at 572~GHz were performed by  the Kuiper Airborne Observatory \citep{Keene} and the Odin satellite \citep[e.g.,][]{Larsson, Persson, wirstrom}. The launch of ESA's 3.5~m Cassegrain telescope, the \emph{Herschel} Space Observatory\footnote{\url{http://herschel.esac.esa.int/}} 
\citep{herschel} opened up unique opportunities to observe rotational transitions 
in the  sub-mm and far-infrared, with high sensitivity and spectral resolution 
using the Heterodyne Instrument for the Far-Infrared  
\citep[HIFI; ][]{HIFI}, e.g., \cite{Hily-Blant, Persson12, Biver, Moscadelli}.
From 2010 and onwards, observations of THz transitions  have also been possible 
using the Stratospheric Observatory For Infrared Astronomy\footnote{\url{http://www.sofia.usra.edu/}} (SOFIA). 

 \begin{figure*}[\!htb] 
\begin{subfigure}{0.5\textwidth}
	\centering
	\includegraphics[width=1.0\linewidth]{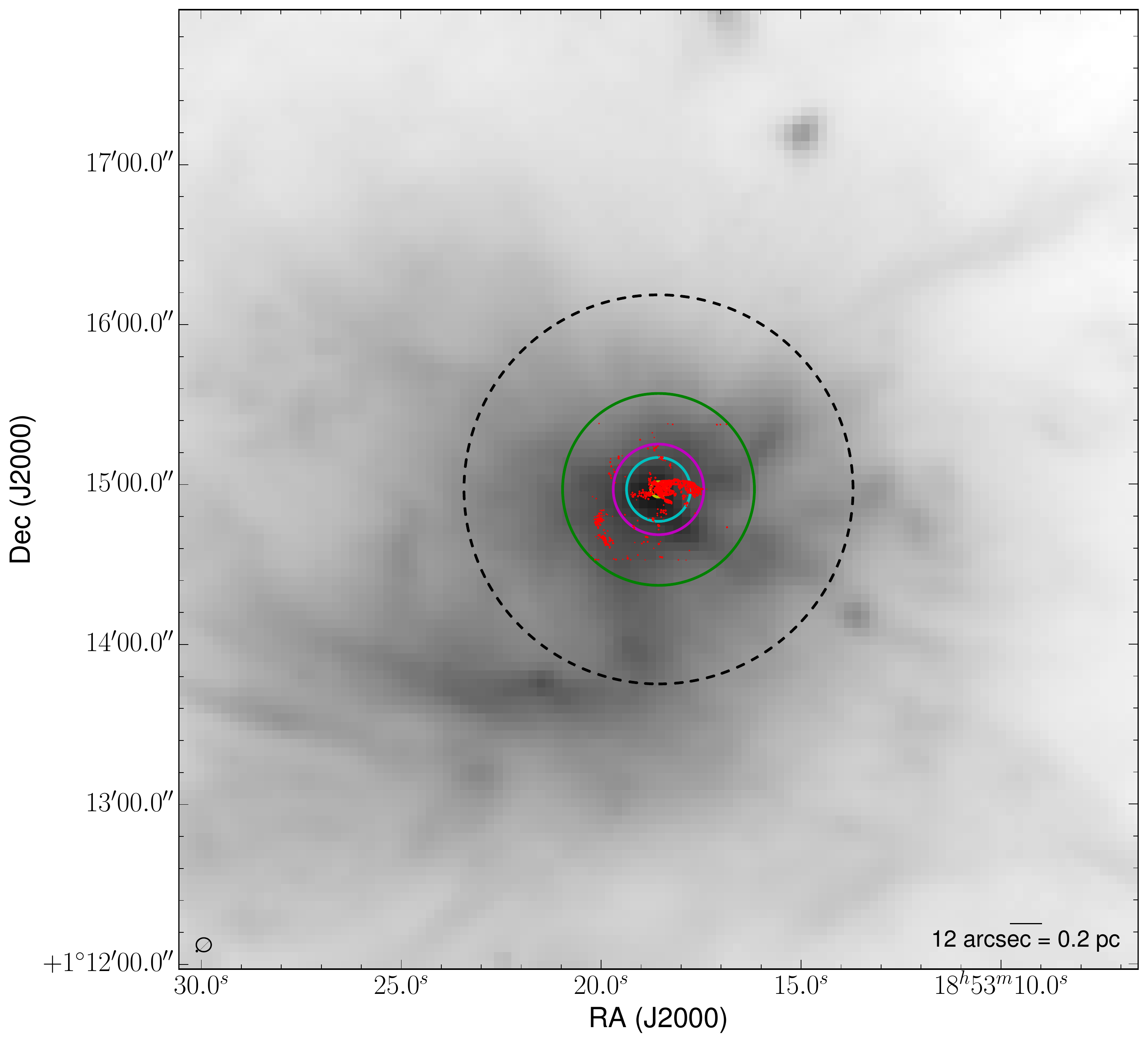}  
	\caption{ 
	}
	\label{Fig:beamsizesA}
\end{subfigure}%
\begin{subfigure}{0.5\textwidth}
	\centering
	\includegraphics[width=1.0\linewidth]{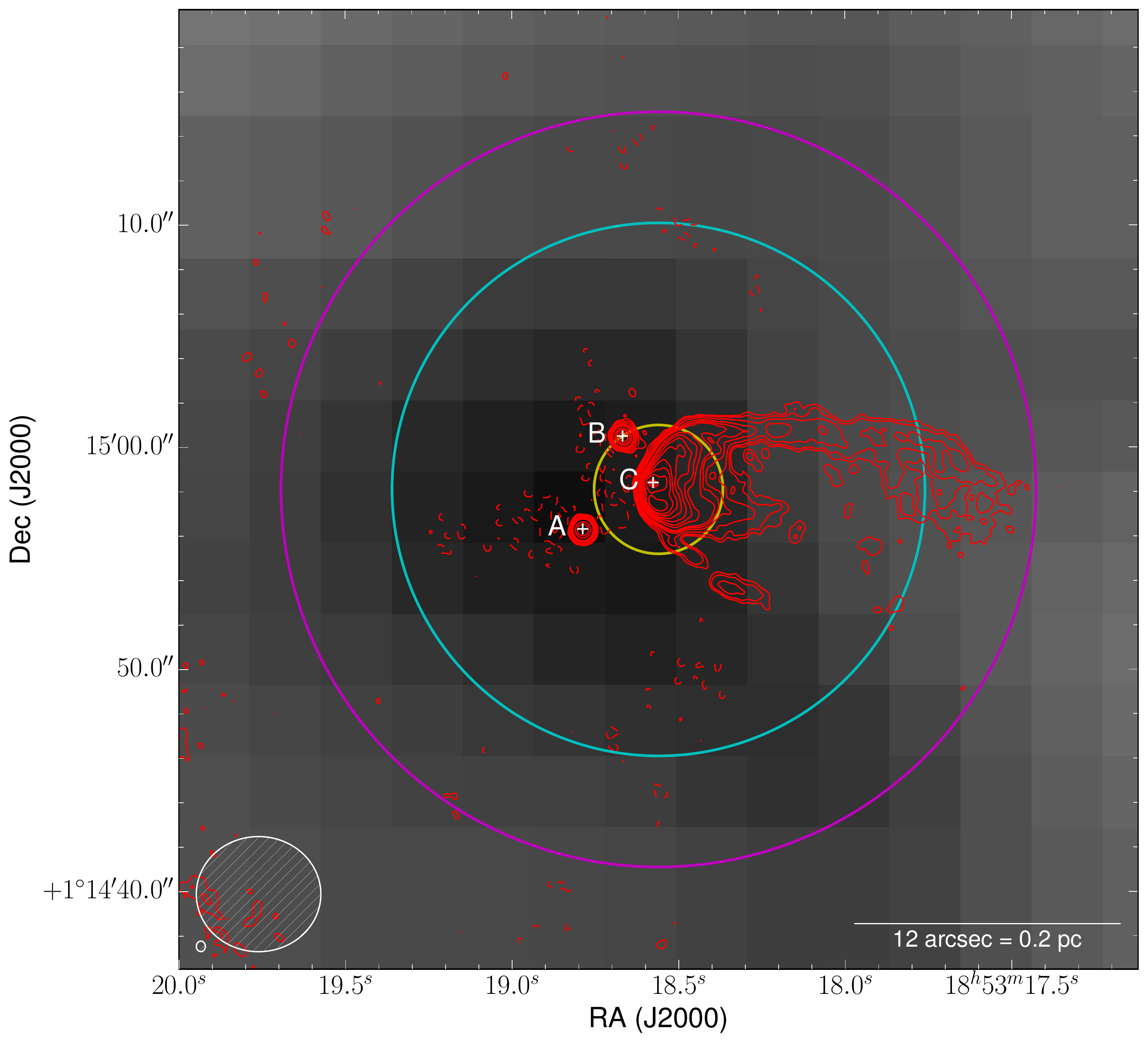}  
	\caption{ 
	}
	\label{Fig:beamsizesB}
\end{subfigure}
\caption{Far-IR grayscale image in both (a) and (b) is observed with the PACS \citep[Photodetector Array Camera and Spectrometer, ][]{PACS} instrument at 70~$\mu$m obtained with \emph{Herschel} as part of the HOBYS
Guaranteed Time key program \citep[PI:\emph{KPGT\_fmotte\_1},][]{HOBYS}. It is overlaid with the contours (\textit{red}) of the 2~cm free-free continuum emission observed by \cite{Sewilo2004}. For the $\lambda$~=~2 cm continuum image the contour levels are at -3, 3, 5, 10, 20, 30, 40, 80, 120, 180, 250, 350, 450, 600, 800, and 850 times the noise level of 0.3 mJy~beam$^{-1}$. 
The pixel scale of the \emph{Herschel}-PACS image is 3.2$\arcsec$. The beam size of $5.2\arcsec\times5.6\arcsec$ for the 70~$\mu$m image and $0.48\arcsec\times0.41\arcsec$ for the 2~cm contour emission are illustrated in the lower left corner. 
The green, magenta and cyan circles illustrate the 36$\arcsec$, 17$\arcsec$ and 12$\arcsec$ Herschel half power beam width, at the respective frequencies 572~GHz, 1214~GHz and 1764~GHz. 
The circles in dashed black and yellow illustrate the model radii $R_{\rm max}$ and $R_{\rm ph}$, respectively. 
}
\label{Fig:beamsizes}
\end{figure*} 
%
 
We present \emph{Herschel}-HIFI observations \textnormal{and modeling results} of seven spectrally resolved ammonia transitions at THz frequencies toward the MSF region G34.26+0.15 (also known as IRAS 18507+0110, hereafter G34), in order to investigate the accretion process. 
These seven transitions have a high range of critical densities which made it possible for us to probe the kinematics of the different layers of G34.
The source, whose properties are summarized in Table~\ref{Table:sample}, is a well-studied hot core with an adjacent cluster of H$_2$O masers \citep{hiroshi, Fey, Benson} and OH masers \citep{Gaume, Garay1985}, \textnormal{indicating the existence of an outflow at subparsec-scale \citep{klaassen}.}
\textnormal{In radio continuum emission, shown in Fig.~\ref{Fig:beamsizes}, the source}  exhibits two very condensed UC \ion{H}{II} regions called A and B, a more evolved compact \ion{H}{II} region with a cometary shape named component C and an extended ringlike \ion{H}{II} region called component D \citep{Mookerjea, Fey}. Component C contains a UC \ion{H}{II} region "head" (at which our observations were centered) and a diffuse "tail" pointing (from tail through head) in the direction of the supernova remnant W44 \citep{Reid}. 
Interferometry at arcsecond resolution of molecular spectral lines, e.g., HCOOCH$_3$ and CH$_3$CH$_2$CN, \textnormal{shows} that the molecular hot core emission partially overlaps with the cm-wavelength continuum emission of component C, and not with that from A and B \citep{Mookerjea}. Radio observations of NH$_3$ \textit{absorption}  locate the hot core between G34.26+0.15 and the observer \citep{Heaton89}.
Infall signatures of the G34 envelope have been found by \cite{wyrowski2012} using THz observations of one ammonia transition performed with SOFIA, \textnormal{and modeling it using RATRAN\footnote{\textnormal{\url{http://www.sron.rug.nl/~vdtak/ratran/frames.html}}}. 
Our modeling, in contrast, is based upon the radiative transfer code ALI (see Sect.~\ref{section:ali}) and simultaneously reproduces seven ammonia line profiles and adjacent continuum fluxes. }

  \begin{table}[\!tb] 
\centering
\caption{Properties of the high-mass star forming region G34.26+0.15.
}
\begin{tabular} {lcccc} 
 \hline\hline
     \noalign{\smallskip}
 $L_{\rm bol}$	& $d$  &	$V_{\rm sys}$	&	$M_{\rm envelope}$	& $R_{\rm envelope}$ \\ 
\noalign{\smallskip}
($L_{\odot}$) 	& (kpc) &	(km$\,$s$^{-1}$)	&	(M$_{\odot}$) & (AU)\\
 \noalign{\smallskip}
     \hline
\noalign{\smallskip}  

$1.9\,\times\,10^{5}$\tablefootmark{$a$}	&  3.3\tablefootmark{$a$}	&	58.1\tablefootmark{$b$} &	1800\tablefootmark{$a$} & $8.1\, \times \,10^{4}$\tablefootmark{$a$} \\
 \noalign{\smallskip} \noalign{\smallskip}
\hline 
\label{Table:sample}
\end{tabular}
\tablefoot{
\tablefoottext{$a$}{\cite{distance3_3kpc}}, \tablefoottext{$b$}{From C$^{17}$O (3--2) line observations with the APEX telescope \citep{wyrowski2012}}.
}
\end{table} 

We present the observations and data reduction in Sect.~\ref{section:obsNred} 
followed by results and analysis in Sects.~\ref{section:spectralproperties} and~\ref{Sect: Analysis}, respectively. 
The radiative transfer modeling is presented in Sect.~\ref{section:ali}. \textnormal{Our modeling results are presented in Sect.~\ref{Section:Result}.} We discuss the results in  Sect.~\ref{section:discussion} and end the paper with  conclusions in Sect.~\ref{section:conclusion}.

\section{Observations and data reduction}
\label{section:obsNred}
  \begin{table*}[\!hbt] 
\centering
\caption{Observed  ammonia transitions toward G34.3$+$0.15 with \emph{Herschel}-HIFI.
}
\begin{tabular} {lccccccccccc} 
 \hline\hline
     \noalign{\smallskip}
& Transition	& Frequency\tablefootmark{$a$}&	$E_{\mathrm{l}}$\tablefootmark{$b$} &	$E_{\mathrm{u}}$\tablefootmark{$c$}	&	$A_{\rm ul}$\tablefootmark{$d$}	&  $n_\mathrm{crit}$\tablefootmark{$e$}    &	$g_{\rm l}/g_{\rm u}$\tablefootmark{$f$} &	HPBW\tablefootmark{$g$}	& 	 $\eta_{\mathrm{mb}}$\tablefootmark{$h$}	&	$T_\mathrm{C}$\tablefootmark{$i$}		&	$1\sigma$/$T_\mathrm{C}$\tablefootmark{$j$}  
\\    \noalign{\smallskip}
& 	($J_{K,\epsilon}$) 	&(GHz)	&	(K)&	(K)	&  $\log_{10}(\rm s^{-1}$)	& (cm$^{-3}$)	& &	($\arcsec$)	 	&	&	(K)	&	(\%)      \\
     \noalign{\smallskip}
     \hline
\noalign{\smallskip}  
o-NH$_3$&   1$_{0\,+}$ -- 0$_{0\,+}$  & 572.498 & 0	& 27	& $-$2.802& 3.4e7	&4/12	&	36 	  & 0.62   &  0.86		& 3.0
\\
&   2$_{0\,+}$ -- 1$_{0\,+}$ & 1214.853  	&27& 86	&	$-$1.742& 3.1e8	&	12/20	&	17	 & 0.61 &   4.35   	& 2.4   
\\
&   3$_{0\,+}$ -- $2_{0\,+}$ &1763.524     	& 86 & 170&	$-$1.226&	  1.1e9	&20/28	&	12	& 0.60   &   6.60   	& 1.1   
\\     
\noalign{\smallskip}
     \hline
     \noalign{\smallskip}
p-NH$_3$&  2$_{1\,-}$ -- 1$_{1\,-}$  &1168.452	&23	& 79	&	$-$1.918&3.1e8	&	6/10	&	18	& 0.61   &   4.50  	& 1.5   
\\
& 2$_{1\,+}$ -- 1$_{1\,+}$   	 & 1215.246&   22&  80	&  	$-$1.867& 3.5e8	&	6/10	&	17	  & 0.61 &   4.37  	& 2.4
\\
&   3$_{1\,+}$ -- 2$_{1\,+}$  	 &1763.601	&80	&165	&	$-$1.277& 1.2e9	&	10/14	&	12	  & 0.60   &   6.60   	& 1.1   
\\						
&   3$_{2\,+}$ -- 2$_{2\,+}$  	 &1763.823	&64	&149	&	$-$1.482& 9.3e8	&	10/14	&	12	  & 0.60   &   6.60   	& 1.1   
\\
    \noalign{\smallskip} \noalign{\smallskip}
\hline 

\label{Table: transitions}
\end{tabular}
\tablefoot{ 
{All spectroscopic data are taken from JPL.}
 \tablefoottext{$a$}{The highest error in frequency  is 0.1~MHz for the 1764 GHz lines.} 
\tablefoottext{$b$}{Lower state energies.} 
\tablefoottext{$c$}{Upper state energies.} 
\tablefoottext{$d$}{The Einstein coefficient for spontaneous emission  $A_{\rm ul}$.}
\tablefoottext{$e$}{Critical density,  $A_{\rm ul}/C_{\rm ul}$, where $C_{\rm ul}$ is the collision coefficient  evaluated for 50~K.}
\tablefoottext{$f$}{Ratio of lower and upper state degeneracies}.
\tablefoottext{$g$}{The \emph{Herschel}  half power beam width at respective frequency \citep{OlbergHIFI14} 
}.
\tablefoottext{$h$}{The main beam efficiency \citep{OlbergHIFI14}.} 
\tablefoottext{$i$}{The observed SSB continuum intensity not corrected for main beam efficiency.}
\tablefoottext{$j$}{The rms noise  relative to $T_\mathrm{C}$}.
}
\end{table*} 
The ground state rotational transitions of ortho-\element[][][][3]{NH}, \mbox{$J_K = 1_{0}-0_{0}$} 
at 572~GHz,   and para-\element[][][][3]{NH}, \mbox{$J_K = 2_{1+}-1_{1+}$} at 1215~GHz, together with 
ortho-\element[][][][3]{NH} \mbox{$J_K = 2_{0}-1_{0}$} at 1214~GHz in the same band,  were observed  
toward G34 with  \textit{Herschel}-HIFI in March--April 2011 as a part of the 
PRISMAS\footnote{\url{http://astro.ens.fr/~PRISMAS}} Guaranteed Time key program 
(PRobing InterStellar Molecules with Absorption line Studies). As a complement to these 
observations, we have observed  
four  excited transitions of both ortho  and para symmetries in April 2012  as part of our 
OT1 program ''Investigation of the nitrogen chemistry in diffuse and dense interstellar gas''. 
All observed transitions are found in Table~\ref{Table: transitions} and  the \emph{Herschel} 
observational identifications are listed 
in Table~\ref{Table:obsid} (on-line material).  The pointing for G34 was centered at 
$\alpha_{J2000}$\,=\,18$^\mathrm{h}$53$^\mathrm{m}$18\fs6, $\delta_{J2000} = +1^\circ$\,14$\arcmin$\,58\farcs0.

We used the Dual Beam Switch mode\footnote{\textnormal{using two OFF positions $3\arcmin$ away from the ON position, and on opposite sides}} and the Wide Band Spectrometer with a 
bandwidth of 2.4~GHz for the highest transitions around 1764~GHz, and 4~GHz in all other bands. An effective spectral resolution of 1.1~MHz was used, corresponding 
to a velocity resolution of 0.2~km$\,$s$^{-1}$ for the 1764~GHz transitions, 
and 0.6~km$\,$s$^{-1}$  at 572~GHz. 
The signal was measured in horizontal and vertical polarizations, which agreed in calibrated intensity within 10~\%.  
Because HIFI is a double side band instrument (DSB),   all observations were performed with 
three different overlaping settings of the local oscillator (LO)
 to  determine the sideband origin of the lines.

 \begin{figure}
\centering
\resizebox{0.86\hsize}{!}{ 
\includegraphics{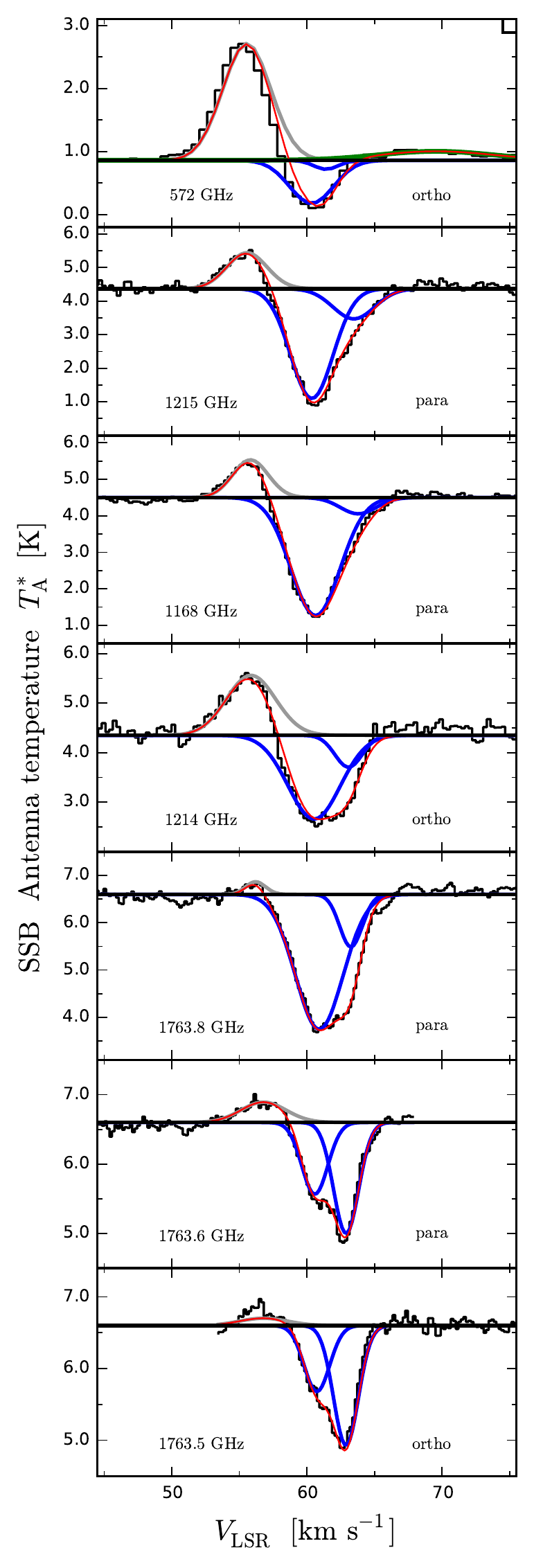}  
}
\caption{Single sideband spectra of all observed \element[][][][3]{NH} transitions 
over the LSR velocity range 45 to 75~km$\,$s$^{-1}$. The horizontal solid line 
represents the continuum level and the colored curves represent Gaussian fits presented in detail in Table~\ref{Table:Gaussian}.
}
\label{Fig:HIPEG34}
\end{figure} 
 
The total calibration uncertainties are \mbox{$\lesssim\,$$8\,$}\% for band~1, 
\mbox{$\lesssim\,$$9\,$}\% for band~5, and \mbox{$\lesssim\,$$14\,$}\% for band~7, 
including the uncertainty in sideband gain ratio. All error components are added in quadrature. 
Detailed information about  the HIFI calibration including errors, beam efficiency, mixer 
sideband ratio, pointing, etc. can be  found  on the Herschel internet
 site\footnote{\url{http://herschel.esac.esa.int/twiki/pub/Public/HifiCalibrationWeb/HifiBeamReleaseNote_Sep2014.pdf}}.
The in-flight performance is  described by \citet{beameff}.

The data were processed using the \textit{Herschel} Interactive Processing 
Environment \cite[HIPE;][]{HIPE}, version 11.0 up to level~2   providing fully calibrated DSB spectra 
on a $T_\mathrm{A}^*$ antenna temperature intensity scale where the lines are calibrated to single sideband (SSB) and the continuum to DSB. We have accordingly divided  the   observed DSB continuum  by two in order to be properly scaled to single sideband (SSB).
The pipeline allows usage of the latest calibration and offers an alternative bandpass 
calibration method that smoothes optical sinusoidal standing waves occurring in band 1--5 
from the internal load. It also removes standing waves at 92 and 98~MHz, which appear from 
the load measurements in all bands, used for intensity and bandpass calibration. 
The   \emph{FitHifiFringe} task is used to fit and remove residual ripples in the spectra 
using three sine functions in the frequency range 30--460~MHz. Three LO-tunings are in good agreement without any visible contamination from the image sidebands. The resulting data quality is consequently excellent with very low intensity ripples. 

The FITS files are then exported to the spectral line software package 
{\tt xs}\footnote{Developed by Per Bergman at Onsala Space Observatory, Sweden; \url{http://www.chalmers.se/rss/oso-en/observations/data-reduction-software}} 
for further analysis. Similar appearance and noise characteristics are found in all individual spectra and  
therefore we average all tunings and both polarizations to the final  noise-weighted spectra. 
First-order polynomial baselines are fitted and subtracted, and  as a final step the mean 
SSB continuum is added.
All spectra in this paper are  shown in a SSB intensity scale. 
The frequency scale is converted to Doppler velocities relative to the local standard of rest 
$\upsilon_\mathrm{LSR}$ using the line frequencies, listed in Table~\ref{Table: transitions}.

For the above line identification work we used the Jet Propulsion Laboratory, JPL\footnote{\url{http://spec.jpl.nasa.gov}} \citep{JPL}, catalogue
entry for NH$_3$, which currently is based upon the extensive re-analysis of \cite{Yu2010}. A review on ammonia is found in \cite{ho} and an investigation of the hyperfine structure components of the rotational transitions has been performed by \cite{Coudert} and of ortho-\element[][][][3]{NH} at 572~GHz by \cite{cazz}.

\begin{figure}
\resizebox{\hsize}{!}{ 
\includegraphics{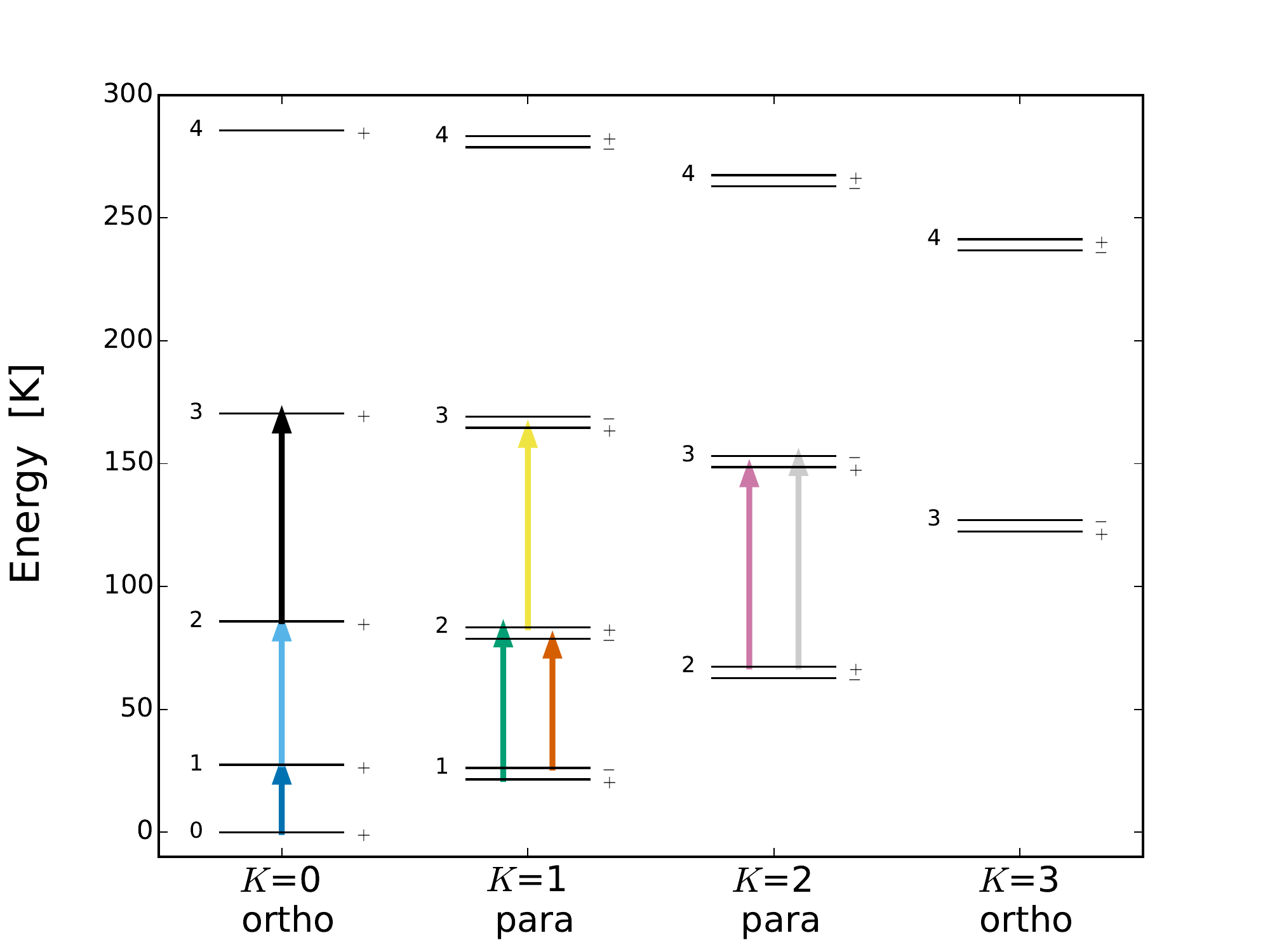} }
\caption{Energy-level diagram of \element[][][][3]{NH},  
where the rotational quantum number, $J$, is placed to the left of each level, the symmetry index, $\epsilon=+$ or $-$, is placed to the right of each level. The symmetry index is that defined by \cite{rist}, which might differ from common spectroscopic notation. The transitions are marked with colored arrows, where the direction corresponds to absorption. \textnormal{The gray arrow represents the 1810~GHz transition observed with SOFIA by \cite{wyrowski2012}.}}
\label{Fig:Energy}
\end{figure}

The observed spectra are shown in Fig.~\ref{Fig:HIPEG34}, while Fig.~\ref{Fig:Energy} shows the energy level diagram of ammonia with its two distinct species, ortho- and para-NH$_{3}$ (see Sect.~\ref{section:OTP}). Note that all observed transitions have non-metastable upper states ($J_u >K$, $K\neq0$), which are easily excited radiatively near luminous IR sources and have short lifetimes for radiative decay. The $J$ levels are further split in a complex hyperfine structure. However, since the   $K$\,=\,0 ladder of energy levels has no inversion splitting,  
the (0,0) ground state can be studied only through the rotational $1_0-0_0$ transition at 572~GHz.  

\section{Spectral properties}
\label{section:spectralproperties}
All transitions show  an inverse P-Cygni profile dominated 
by absorption at $V_{\rm LSR}\approx 58-66$~km$\,$s$^{-1}$, except for the fundamental $1_0-0_0$ ortho transition at 572~GHz, which has a strong emission of 1.85~K and an almost saturated   absorption. 
The 572~GHz line also  shows much less absorption  in the redshifted high velocity wing 
especially compared to the highest excited transitions at 1763~GHz. The two lowest para transitions at 1168 and 1215~GHz, both connecting to the $2_1-1_1$ state, and the $2_0-1_0$ ortho transition at 1214~GHz show similar emission of $\sim$1.2~K, but  with weaker absorption in the ortho line.
The highest excited  $J = 3-2$ transitions around 1763~GHz,  with upper energy levels of 149--170~K, appear mainly in absorption. Note that even though the ortho $3_{0,+}-2_{0,+}$  and  para $3_{1,+}-2_{1,+}$ transitions   are separated by only 77~MHz, corresponding to 13~km$\,$s$^{-1}$, there is no overlap since  the lines show absorption and emission over $\sim$11~km$\,$s$^{-1}$. The lowest  transitions show broader emission and absorption features than the excited  lines. 
The rms noise and single side band (SSB) continuum  levels, $T_\mathrm{C}$, for  
all transitions are listed in Table~\ref{Table: transitions}.

\begin{figure}[\!htb]
 \centering 
\resizebox{\hsize}{!}{ 
\includegraphics{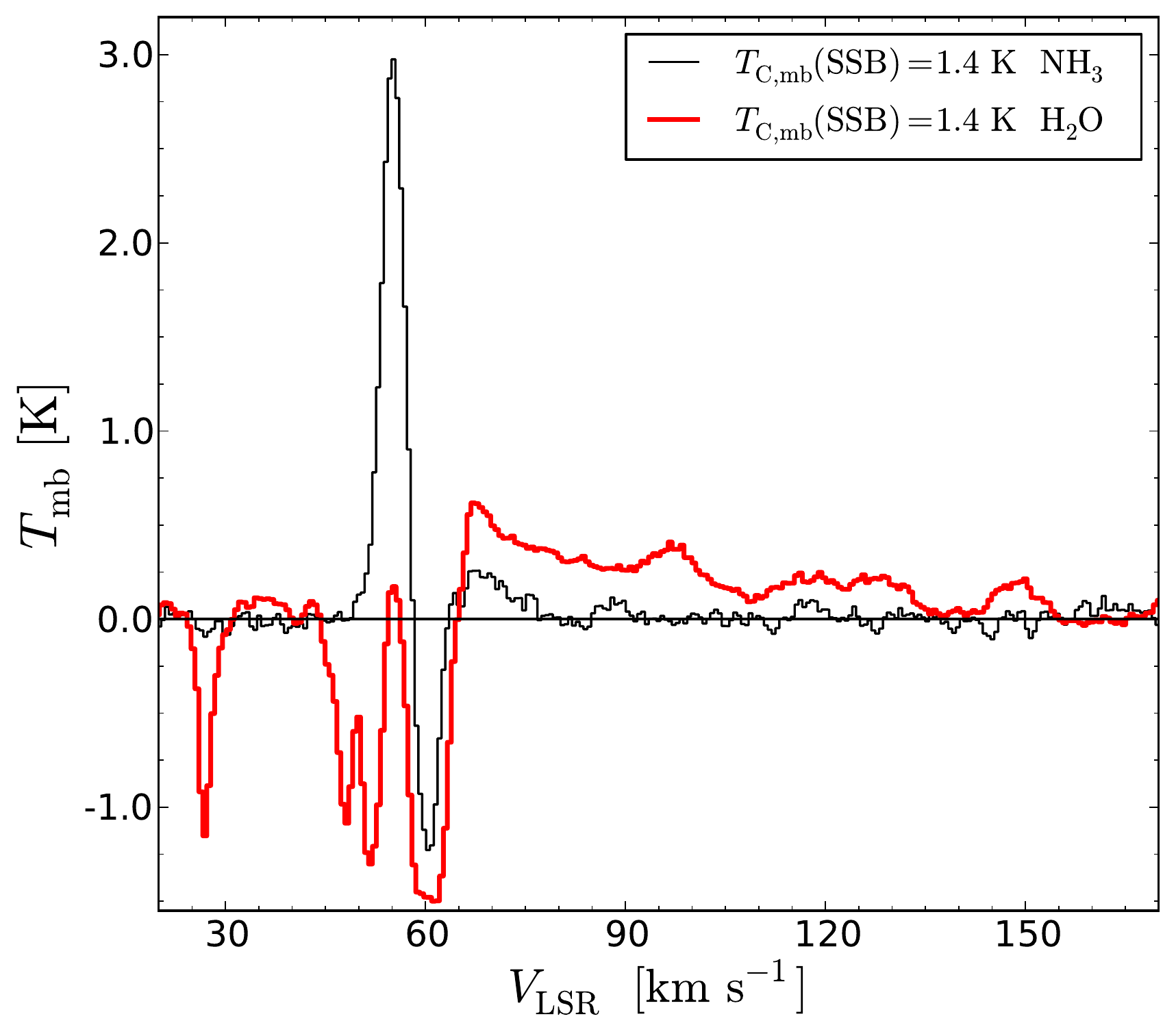}  
}
\caption{Comparison between o-NH$_{3}$ (572~GHz) and o-H$_{2}$O (557~GHz) \textnormal{\emph{Herschel}-HIFI} data from \cite{Flagey}.  The redshifted emission wing of water displays a clear signature of a broad outflow component from the source. 
Note that the water absorption is saturated around the systemic velocity \textnormal{($\approx 62~$km$\,$s$^{-1}$)}.
}
\label{Fig:outflow}
\end{figure}

The ortho 572~GHz line also shows evidence of a weak, broad and redshifted emission component not seen in any other line. 
In Fig.~\ref{Fig:outflow}, we compare the line profiles of the   $1_{1,0}-1_{0,1}$ ortho-H$_2$O \cite[][]{Flagey} and the  $1_0-0_0$  ortho-NH$_{3}$ transitions. Water  shows a very broad emission from the source velocity up to $\sim$$150$~km$\,$s$^{-1}$ whereas ammonia, being the less sensitive to shocks and outflows, shows broad emission extending only to $\sim$$75$~km$\,$s$^{-1}$. 

\begin{figure*}[\!htb] 
 \centering
\resizebox{0.9\hsize}{!}{ 
\includegraphics{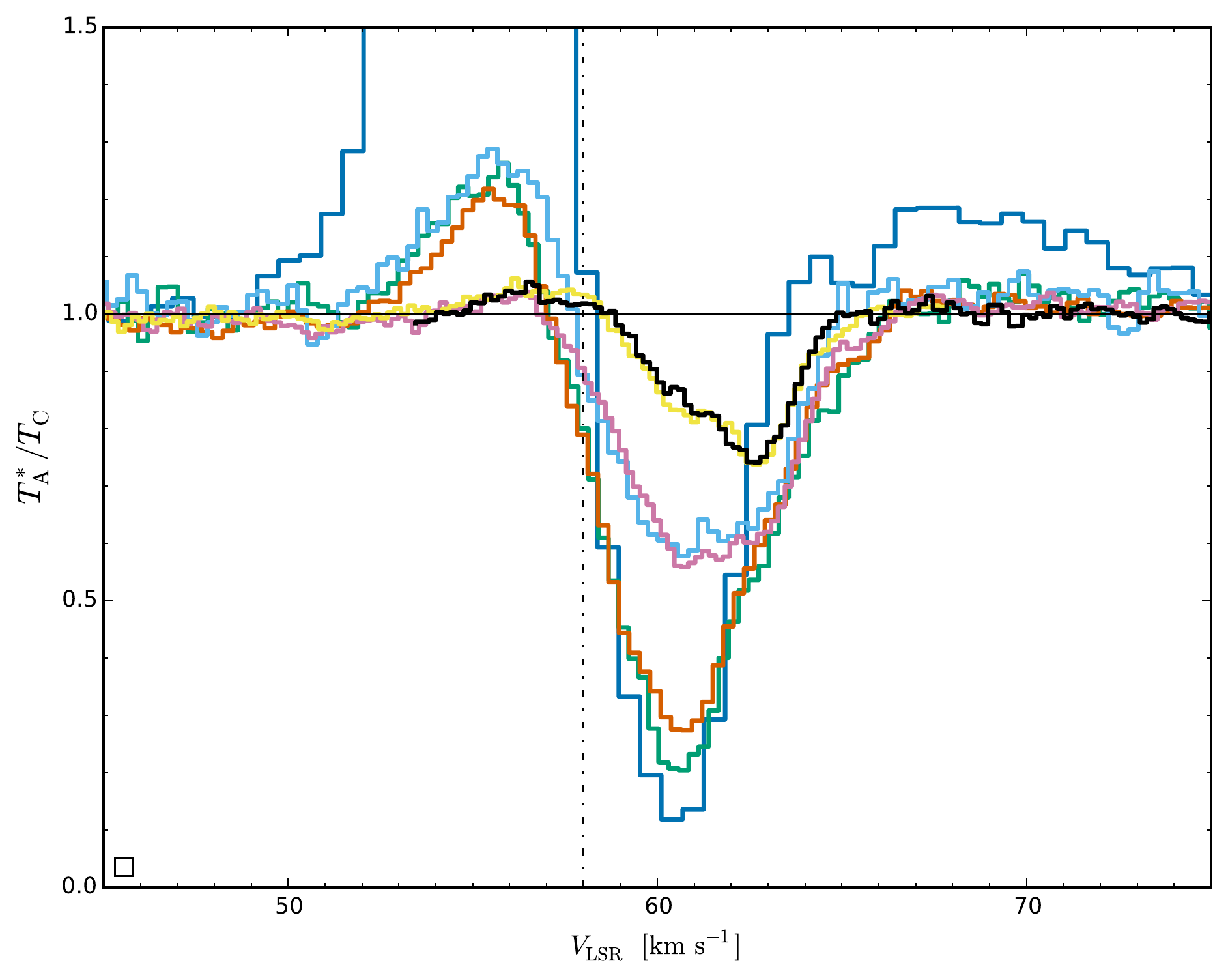}  
}
\caption{Normalized  spectra of all observed \element[][][][3]{NH} transitions 
used in the analysis of the \emph{absorption} line profiles 
(same color code as in Fig.~\ref{Fig:Energy}).  The vertical black dotted line marks the systemic velocity of G34.}
\label{Fig:G34}
\end{figure*}

To allow a comparison of the \emph{absorption} line profiles, all spectra are normalized as $T^{*}_{\rm A}/T_{\rm C}$ and shown in Fig.~\ref{Fig:G34}. 
The velocity at the minimum intensity increases with energy level.
Together with the inverse P~Cygni profiles this suggests that we are tracing gas with inward motion, confirming the results by \cite{wyrowski2012}.   
The different lines can thus be used to trace different layers in the absorbing gas.  
The critical densities are  listed in Table~\ref{Table: transitions} and range 
from $\sim$$10^7$~cm$^{-3}$  for the 572~GHz  transition to $\sim$$10^9$~cm$^{-3}$ for the highest excited  lines. 

We find no evidence of ammonia absorption in diffuse gas along the sight-line toward G34, in contrast to other species such as water, which shows strong foreground absorption features in the velocity range \mbox{$10-55$~km$\,$s$^{-1}$}   \citep{Flagey}.

\section{Analysis}
\label{Sect: Analysis}

An analysis of  the ammonia line profiles is not straight-forward due to the complex blend of emission and absorption features and an excitation most likely not in Local Thermodynamic Equilibrium (LTE). However, in this section we will present estimates of the velocity gradients, opacities, column density, ortho-to-para  ratio and source sizes, which can be used as starting values for the more elaborate modeling in Sect.~\ref{section:ali} using an Accelerated Lambda Iteration code. 

\subsection{The absorbing material}

\subsubsection{Velocity gradients}
\label{Sect:Velocity}
\begin{figure}[!htb] 
  \centering
\resizebox{0.95\hsize}{!}{ 
\includegraphics{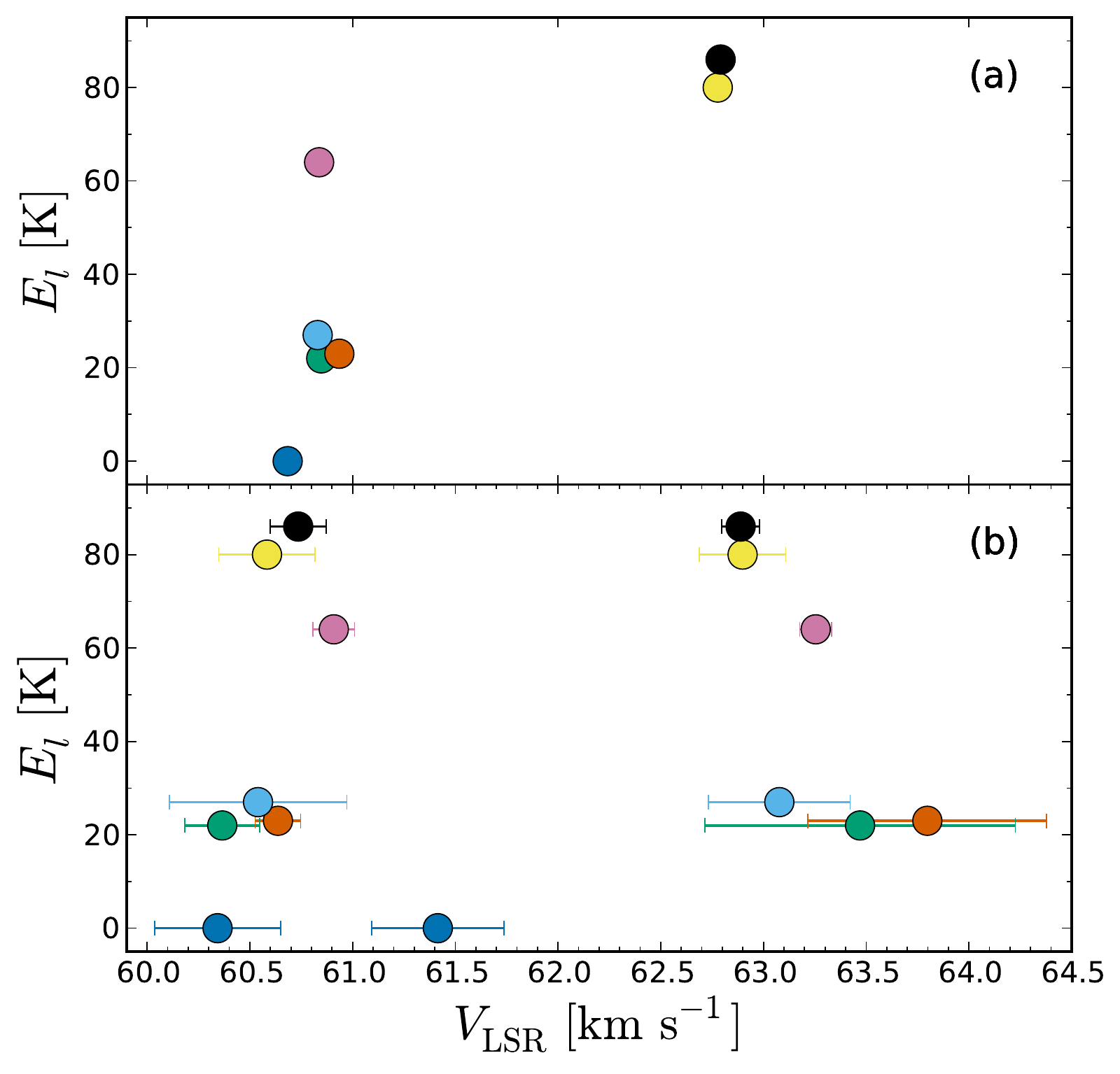}  
}
\caption{Lower state energies  plotted as a function of $V_{\rm LSR}$ (a) at the minimum intensity and (b) at the minimum intensity of the Gaussian fits of the two absorption profiles (see Table~\ref{Table:Gaussian}). The strong emission and weak absorption at higher velocity may explain the low $V_{\rm LSR}$ for the 572 GHz line. Same color code is used as in Fig.~\ref{Fig:Energy}.
}
\label{Fig:non-uniform}
\end{figure}

To quantify the velocity gradient, the lower state energies of  all transitions 
are plotted in Fig.~\ref{Fig:non-uniform} 
as a function of the $V_\mathrm{LSR}$ of the minimum absorption line intensities. 
Comparing the result of minimum intensity measured directly from observed spectra and from gaussian fits of two absorption profiles (Table~\ref{Table:Gaussian}), we find that the lines cluster around two velocities, 
$\sim$60.6 and  $\sim$63.0~km$\,$s$^{-1}$, 
suggesting the existence of  \emph{two} gas components moving inwards toward the center of the cloud, each with approximately  constant velocity. 
The lines of highest excitation, 1763.5~GHz (ortho) and 1763.6~GHz (para), have negligible emission and thus trace the absorption velocities rather accurately with a minimum intensity at $\sim$63.0~km$\,$s$^{-1}$ although they also display absorption at lower velocities.
The lines of intermediate excitation, 1214 (ortho) and 1763.8~GHz (para), show approximately equal absorption in both velocity components, whereas the lowest ortho- and para-lines at 572, 1168 and 1215~GHz mainly show absorption at  $\sim$60.6~km$\,$s$^{-1}$.
The velocities at the minimum intensity in these lines are
more uncertain due to their stronger emission, in particular the 572~GHz line. However, since the 1168, 1215 and 1214~GHz lines have very similar emission over $V_\mathrm{LSR}\approx51-58$~km$\,$s$^{-1}$, and the 1214 and 1763.8~GHz lines also have almost identical absorption line profiles over $V_\mathrm{LSR}\approx58-66$~km$\,$s$^{-1}$ (with emission only in the 1214~GHz line), the peaks of emission and absorption are sufficiently well separated that the absorption velocities are well determined.

 \subsubsection{Optical depth and column density}
\label{Sect: column}
In order to parameterize the spectral lines, we have performed multicomponent Gaussian fits to all lines (see Fig.\ref{Fig:HIPEG34}). We list the results in Table~\ref{Table:Gaussian} and display the fits in Fig.~\ref{Fig:HIPEG34}. For the 572~GHz transition, an additional component corresponding to the broad emission line was included.

 \begin{table*}
\centering
\caption{Results from Gaussian fits, $T_{\rm l}$,  $\Delta V$, and  $V_{\mathrm{LSR}}$ for the emission, 
calculated peak opacities, $\tau_\mathrm{p}$, integrated opacities, $\int\tau_\nu d V$, and 
column densities of the absorbing material in respective lower state, $N_\mathrm{l}$.}
\begin{tabular} {lccccccc} 
 \hline\hline
     \noalign{\smallskip}
&Freq 	&  $T_{\rm l}$\tablefootmark{$a$}	& $\Delta V$\tablefootmark{$b$} & $V_{\mathrm{LSR}}$  & $\tau_\mathrm{p}$\tablefootmark{$c$} &
$\int\tau_\nu d V$\tablefootmark{$c$}   	& $N_{\rm l}$\tablefootmark{$d$}   
\\ \noalign{\smallskip}
&(GHz) &  (K) & (km$\,$s$^{-1}$) &  (km$\,$s$^{-1}$)&  & (km$\,$s$^{-1}$)	& ($\rm cm^{-2}$) 
\\    
 \noalign{\smallskip}
     \hline
\noalign{\smallskip}  

o-NH$_3$& 572.498 & $-0.7\pm0.1$		&  $3.8\pm0.3$ 	&	$60.3\pm0.3$ 		&  1.3	&	5.4  & 	2.0e13	\\
&		& $-0.1\pm0.1$		&  $2.1\pm0.9$		&	$61.4\pm0.3$ 		&  0.2	&	$\lesssim$0.4  & 	$\lesssim$1.5e12	\\
&		& $1.85\pm0.01$	&  $4.2\pm0.07$  	&	$55.55\pm0.04$ & 	&	... & 		... 	\\
&		& $0.14\pm0.01$	&  $10.0\pm0.8$ 	&	$69.5\pm0.3$ &	&	... & 		...  
\\  \noalign{\smallskip}
&1214.853  	& $-1.7\pm0.1$	&  $4.5\pm0.9$  &	$60.5\pm0.4$	&  0.5 	&	2.3 & 	1.3e13	 \\ 
&		& $-0.6\pm0.4$		&  $2.4\pm0.7$		&	$63.1\pm0.3$	&  0.2 	&	$\lesssim$0.4 & 	$\lesssim$2.2e12	 \\ 
&		& $1.2\pm0.1$	&  $4.3\pm0.4$  	&	$55.9\pm0.3$	& 	&	... & 		...	  
\\  \noalign{\smallskip}
&1763.524  	& $-0.92\pm0.2$	&  $2.2\pm0.5$	&	$60.8\pm0.1$ 	&  0.2	&	0.4 &		2.2e12  \\ 
&		& $-1.7\pm0.1$	&  $2.2\pm0.1$		&	$62.8\pm0.1$	&  0.3	&	0.7 &		4.1e12  \\ 
&		& $0.10\pm0.08$	&  $4.2\pm18$  	&	$56.8\pm6.7$   & 	&	... & ...	
\\ \noalign{\smallskip} \hline \noalign{\smallskip}
p-NH$_3$ & 1168.452 	& $-3.2\pm0.1$		&  $4.3\pm0.3$		&	$60.6\pm0.1$		& 1.3		&	5.7 & 4.2e13  \\
&		& $-0.4\pm0.3$		&  $3.1\pm0.7$		&	$63.8\pm0.6$		& 0.1		&	$\lesssim$0.3 & $\lesssim$2.6e12  \\
&		& $1.	03\pm0.04$	&  $3.1\pm0.2$  	&	$55.8\pm0.1$ & 	&	... & ...	  
\\  \noalign{\smallskip}
&1215.246 & $-3.3\pm0.3$		&  $4.3\pm0.3$ &	$60.4\pm0.2$		& 1.4	&	5.4 & 	4.0e13	\\
&		& $-0.9\pm0.6$		&  $3.1\pm0.7$  	&	$63.5\pm0.8$		& 0.2	&	$\lesssim$0.9 & 	$\lesssim$6.6e12	\\
&		& $1.07\pm0.04$	&  $3.5\pm0.3$   	&	$55.8\pm0.1$	&	&	... & 		...	
\\  \noalign{\smallskip}
&1763.601& $-1.04\pm0.05$	&  $2.3\pm0.2$ 	&	$60.6\pm0.2$  	& 0.2	&	0.4 & 	2.8e12  \\ 
&		& $-1.60\pm0.05$	&  $2.2\pm0.1$  	&	$62.9\pm0.2$  	& 0.3	&	0.6 & 	4.4e12  \\ 
&		& $0.	29\pm0.02$		&  $3.9\pm0.8$  	&	$56.8\pm0.2$ 		& 	&	... & ...	
\\  \noalign{\smallskip}
&1763.823  	& $-2.85\pm0.04$	&  $4.1\pm0.2$		&	$60.9\pm0.1$		& 0.6	&	 2.5 &	2.7e13    \\ 
&		& $-1.1\pm0.2$		&  $2.0\pm0.2$		&	$63.3\pm$0.1		& 0.2	&	 0.4 &	4.2e12    \\ 
&		& $0.27\pm0.04$		&  $1.7\pm0.4$  	&	$56.2\pm0.1$ & 	&	... & ...	
\\  \noalign{\smallskip}
\hline 
\label{Table:Gaussian}
\end{tabular}
\tablefoot{
\tablefoottext{$a$}{Amplitude of Gaussian fit. A negative value  represents absorption and
a positive value  emission.}
\tablefoottext{$b$}{Line width from Gaussian fit.}
\tablefoottext{$c$}{Estimated with Eq.~(\ref{opacity}).}
\tablefoottext{$d$}{Estimated with Eq.~(\ref{Eq:Nl}), and listed as upper limits for components that are not statistically significant in the Gaussian fitting.}
}
\end{table*} 

We calculate the integrated optical depth for the absorption components  obtained from the Gaussian fits as 
\begin{equation}\label{opacity}
\int\tau_{\nu} \, \mathrm{d} V \approx 1.06    \,\tau_\mathrm{p}\,\Delta V
\approx - 1.06  \, {\rm ln}\, 
\bigg(1  - \frac{|T_{\rm l}| )}{T_{\rm C}}\bigg)\,\Delta V \ ,
\end{equation}
where $\tau_\mathrm{p}$ is the peak opacity, $\Delta V$ is the line width, $|T_{\rm l}|$ is the absolute value of the amplitude 
of the Gaussian fit, and $T_{\rm C}$ is the sum of the continuum and the emission contribution to the continuum, $T_{\rm C}({\rm EM})$, which is estimated using the gaussian emission profile (Table~\ref{Table:Gaussian}).
Assuming that ammonia and water co-exist, we use the fact that the water line  shows saturated absorption at the zero level  to support our underlying assumption in Eq.~\ref{opacity} that the absorbing material completely covers the background continuum within the beam. 
The ammonia line profiles exhibit a mixture of emission and absorption in the presence of a strong submm-wave continuum. Such spectra suggest that the radiating envelope is stratified and that it contains gaseous molecules like \element[][][][3]{NH} mixed with dust. The continuum radiation is intense; for example, $T_{\rm b} = 4.4$~K at 
1215~GHz corresponds to a surface brightness $I_{\nu} = 3.3\times10^{-15}$~W~m$^{-2}$~Hz$^{-1}$~sr$^{-1}$ 
averaged over the 17$\arcsec$ \emph{Herschel} beam. This means that the absorption rate in the $2_{1\,+}\gets 1_{1\,+}$ transition is
$$ I_{\nu} B_{\rm \ell,u} = 2.4\times10^{-3} \;\;\;[{\rm s}^{-1}]\;, $$
where $B_{\rm \ell,u}$ is the Einstein coefficient for absorption. In comparison with the rate coefficient for collisional excitation in the same transition at a collision temperature of 50 K,  \mbox{$C_{21}=4.0\times\,10^{-11}$}~cm$^3$~s$^{-1}$ \citep{danby}, a hydrogen density of \mbox{$n({\rm H}_2) =  6.0\times10^{7}$~cm$^{-3}$}  would be required for collisions to compete with radiative processes (cf. the critical density for emission in Table~\ref{Table: transitions}). Consequently, the excitation of \element[][][][3]{NH} is probably not in LTE and thus requires analysis by means of a non-LTE radiative transfer model of a dynamical envelope.   

Although the continuum radiation of G34 is non-negligible for radiative excitation in the full non-LTE treatment of NH$_{3}$, the resulting excitation temperatures do not significantly change the results of the simple analysis.  
We have here neglected the contribution of emission assuming that 
\mbox{$J(T_\mathrm{ex})\,$=$\,h\,\nu/k \times (\exp(h\,\nu/(k\,T_{\rm ex}))-1)^{-1} \ll  T_\mathrm{C}$},  
and that the cosmic microwave background (CMB), is negligible at the 
frequencies considered here (<1.5~mK). 

Using the simple analysis, the column densities of respective lower state can now be calculated with  
\begin{equation}
N_{\rm l}\, = \, 8\pi\frac{g_{\rm l}}{g_{\rm u}} \frac{\nu^{3}}{c^{3}} \frac{\int\tau_\nu d V}{A_{\rm ul}} \frac{1}{1-{\rm exp}(-\frac{h\nu}{kT_{\rm ex}})} \approx 
\, 8\pi\frac{g_{\rm l}}{g_{\rm u}} \frac{\nu^{3}}{c^{3}} \frac{\int\tau_\nu d V}{A_{\rm ul}}\ ,
\label{Eq:Nl}
\end{equation}
where $g_{\rm l}/g_{\rm u}$ is the ratio of the degeneracy factors of the lower and upper levels and $A_{\rm ul}$ is the Einstein coefficient for spontaneous emission. 
The degeneracy factors and the Einstein coefficients are listed in Table~\ref{Table: transitions}.

A lower limit to the total ammonia column density in the absorbing material then is found by 
summing all $J=0$, 1 and 2 lower state column densities listed in Table~\ref{Table:Gaussian} 
\begin{equation*}
N_{\rm tot} = \sum N_{\rm l} \geqslant  2.1\times10^{14}\,\,\, [\rm cm^{-2}]\ .
\end{equation*} 
In the above summation   we have also added the two unobserved para transitions
\mbox{3$_{2\,-} \leftarrow 2_{2\,+}$} (at 1810~GHz) and  \mbox{3$_{1\,+} \leftarrow 2_{1\,-}$} (1809~GHz)
by counting our observed \mbox{3$_{2\,+} \leftarrow 2_{2\,+}$} and  \mbox{3$_{1\,+} \leftarrow 2_{1\,+}$} columns 
twice since their intensities are expected to be similar. 
The  1810~GHz line was observed   by \cite{wyrowski2012}, who confirm our estimated intensity.
The total column density can be higher if higher levels than $J=2$ are populated. At cm wavelengths the (3,3) inversion line can for instance  be rather strong indicating non-negligible population of this level. 
%

%

Neglect of emission in the submm lines causes the total column density to be underestimated by a factor of two,
based on the excitation temperatures (at $r=R_{\rm max}$) computed in our best-fitting non-LTE models (see Sect.~5 and Fig.~\ref{Fig:Tex}) and $\mbox{$\tau = -\ln{(|T_{\rm l}|-J(T_{\rm ex}))/(T_{\rm C} + T_{\rm C}({\rm EM})-J(T_{\rm ex}))}$}$.

\subsection{Size estimates} 
\label{sect:size}

\begin{figure}[\!htb] 
  \centering
\resizebox{\hsize}{!}{ 
\includegraphics{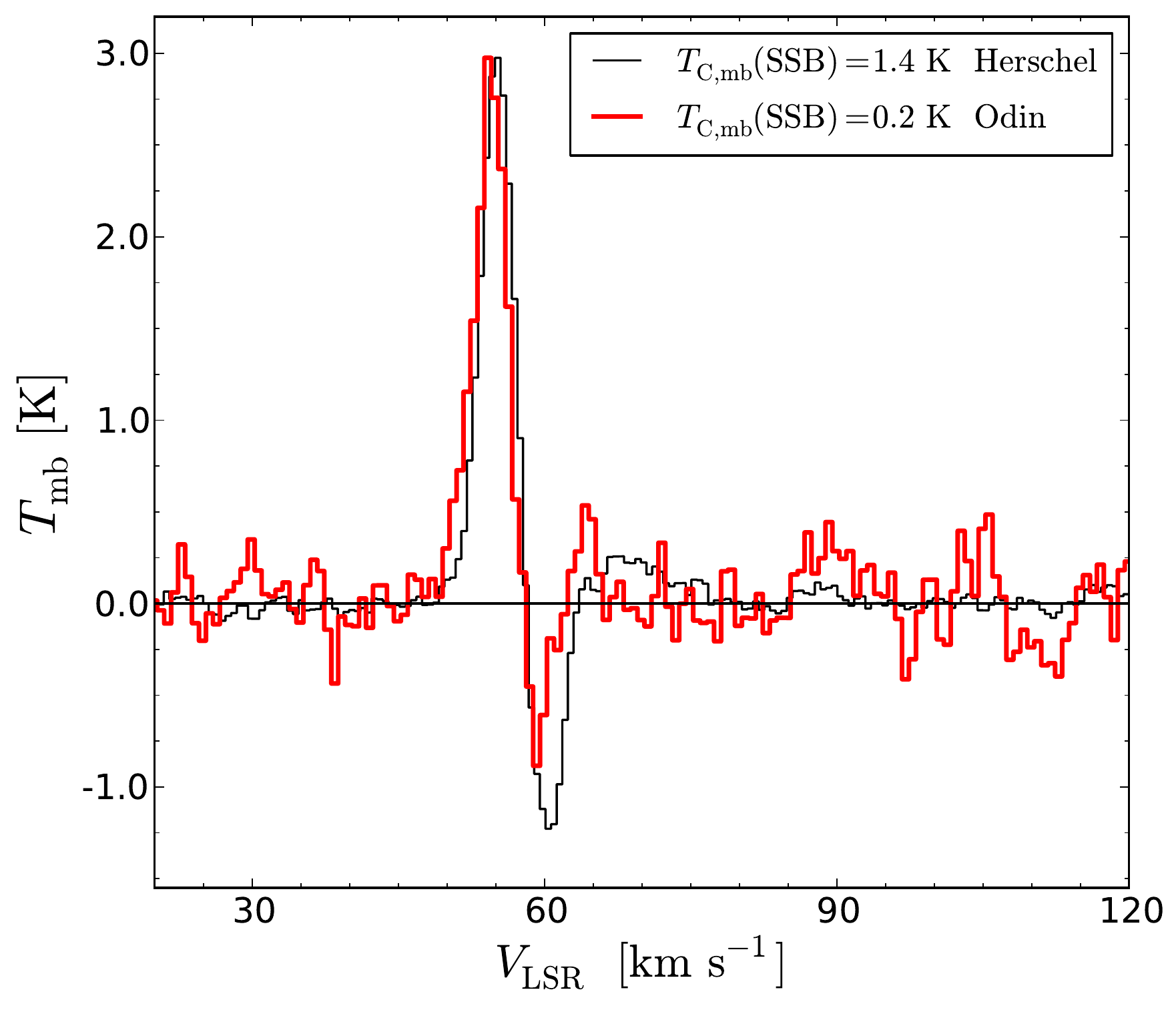}  
}
\caption{Odin   and \emph{Herschel} observations of ortho-\element[][][][3]{NH} 1$_0$ -- 0$_0$ at 572 GHz.
The continuum levels, noted in the legend, are subtracted and the Odin spectra is scaled up with a factor of 7, to allow a comparison of the spectra.
}
\label{Fig:size}
\end{figure}

Our $1_0-0_0$ spectrum can also be compared to observations of this transition performed with the 
Odin satellite \citep{Nordh} toward G34 \citep{odin}. Fig.~\ref{Fig:size} shows the comparison. The ratio of the \emph{Herschel} and Odin antenna temperatures gives an estimate of the sizes 
of the emission   and continuum components according to 
\begin{equation}
\frac{T^{*}_{\rm A,Herschel}}{T^{*}_{\rm A,Odin}}	=	\frac{ \eta_{\rm mb, Herschel}}{\eta_{\rm mb, Odin}} \times \frac{\theta_{\rm s}^{2}+\theta_{\rm mb, Odin}^{2}}{\theta_{\rm s}^{2}+\theta_{\rm mb, Herschel}^{2}}\ , 
\label{Eqs:size}
\end{equation}
where $\theta_\mathrm{s}$ is the effective circular Gaussian source size in arcseconds, 
$\eta_\mathrm{mb, Odin}=0.9$, and $\theta_{\rm mb, Odin}=126$\arcsec, at 572 GHz.

The resulting mean source sizes are comparable to the 36\arcsec \, \emph{Herschel} beam, 
$\sim35\arcsec$ and $\sim38\arcsec$ for the narrow emission and the continuum, respectively. This supports the assertion that it is the dust, mixed with the emitting gas,  that is responsible for the major part of the continuum emission. The size of the  broad   emission  is derived using the Odin noise level  as  1$\sigma$ ($\sim$29~mK) upper limit, which we  find to be $\lesssim34\arcsec$. Small variations of the ratio of antenna temperatures have, however, a high impact on the derived  sizes. A 20\% calibration uncertainty in both spectra gives for instance an error of $\pm20\arcsec$. 
\subsection{Ortho-to-Para ratio} \label{section:OTP}
Two possible relative orientations of the hydrogen spins create two distinct species: ortho-\element[][][][3]{NH} (all spins are parallel, $K$\,=\,3$n$ where $n$ is an integer $\geq 0$) and para-\element[][][][3]{NH} (not all H spins are parallel, $K\neq3n$). Rotational transitions  normally are not allowed between ortho and para states. Neither can radiative or non-reactive collisional transition change the spin orientation between the two symmetries, once formed.
Measurements of the ortho-to-para ratio (OPR) can give valuable insights  
into the competing processes of formation and destruction, radiative and collisional 
excitation, and reactive interchange processes in interstellar space. 
 
Ammonia formation in the gas-phase at high temperatures is 
expected to result in OPR ratios close to
the statistical value of one, because the energy release in the formation process is much higher than the energy difference between the lowest para and ortho states (22~K). If ortho-NH$_{3}$ is formed or condensed at low temperatures (<30~K) on water ice covered dust grains and then desorbed when the grain is heated to 100~K, the OPR may be higher than one.

Our new observations and analysis of three ortho and three para lines probing different energies 
can for the first time be used to estimate the OPR in dense ($10^4$~<~$n_{\rm H_2}$~<~$10^7$~cm$^{-3}$) and warm (15~<~T~<~100~K) gas in star-forming regions.   
Using the results from Table~\ref{Table:Gaussian} we find  
\begin{equation*}
N(\mathrm{ortho}) =\sum N_{\rm l}(\mathrm{ortho}) \geqslant  0.43\times10^{14}\,\,\, [\rm cm^{-2}]\ ,
\end{equation*}
and 
\begin{equation*}
N (\mathrm{para}) =\sum N_{\rm l}(\mathrm{para}) \geqslant  1.68\times10^{14}\,\,\, [\rm cm^{-2}]\ ,
\end{equation*}
that gives us an OPR$\sim$0.3 in the absorbing material, below unity.

This result takes into account all levels of $J<3$ and $K<3$, but not the $J\geqslant3$ levels, which may be populated. Since the ($J,K$)=(3,3) levels would be proportionally most populated, the effect of omitting higher level populations in this OPR estimate makes it effectively a lower limit. However, as will be shown with non-LTE radiative transfer models (below), the computed population in the (3,3) levels account for at most 10\% and $10^{-4}$\% of the total population in the innermost and outermost region, respectively. The inclusion of these levels could no more than double the estimated OPR. 
Thus, the indication of an OPR less than unity is quite secure even considering the effects of excitation. 

\section{Spherical non-LTE models}
\label{section:ali} 
Our basic analysis (in Sect.~\ref{Sect: column}) already suggests that the $T_{\rm ex}$ is different for different transitions, and therefore it is important to perform non-LTE modeling. In order to construct an accurate model that properly accounts for the line overlap and coupling to the continuum, we adopt a non-LTE, one-dimensional, radiative transfer model based on the accelerated lambda iteration \cite[ALI,][]{Hummer91,Hummer92} method. ALI solves the coupled problem of radiative transfer and statistical equilibrium from an initial guess of the level populations. 
The code we employed has been used by \citet{wirstrom}, \citet{PerBj}, and was benchmarked by \cite{maercker}, where the technique is described in more detail. 

\textnormal{Given the complexity of the MSF region, the locations and sizes of the HIFI submm-wave beams are plotted relative to the entire structure in Fig.~\ref{Fig:beamsizes}.  The two smallest HIFI-beams (cyan and magenta) cover the densest part of the complex region emitting at 70~$\mu$m, overlapping component A, B and the head of component C. The largest beam (green) in addition includes the less dense and non-symmetric far-IR region.}
\textnormal{Fig.~\ref{Fig:beamsizes} also illustrates the size of the modeled spherically symmetric source and its inner IR-continuum object. Considering that the model is spherically symmetric and that the region contained within the HIFI submm-wave beams exhibits enormous complexity, it does give a useful representation of the molecular envelope over the scale of the projected beam. }

\textnormal{The physical properties of the spherically symmetric model, like density and temperature, can be varied over a number of homogeneous shells, which move radially at a given velocity (Sect.~\ref{Sect:non-LTE:Procedure}). }
\textnormal{An ensemble of models is constructed for a range of fixed and varying input parameters in a large grid (Sect.~\ref{Sect:Grid}).}
\textnormal{The best-fitting model is identified by minimizing the difference between observed and computed spectral line profiles (Sect.~\ref{sect:chi2}), and by
requiring that the continuum is reproduced to within 10\% (Sect.~\ref{Section:Result}). }

\subsection{Procedure}\label{Sect:non-LTE:Procedure}
\begin{figure}[\!htb] 
 \centering
\resizebox{\hsize}{!}{ 
\includegraphics{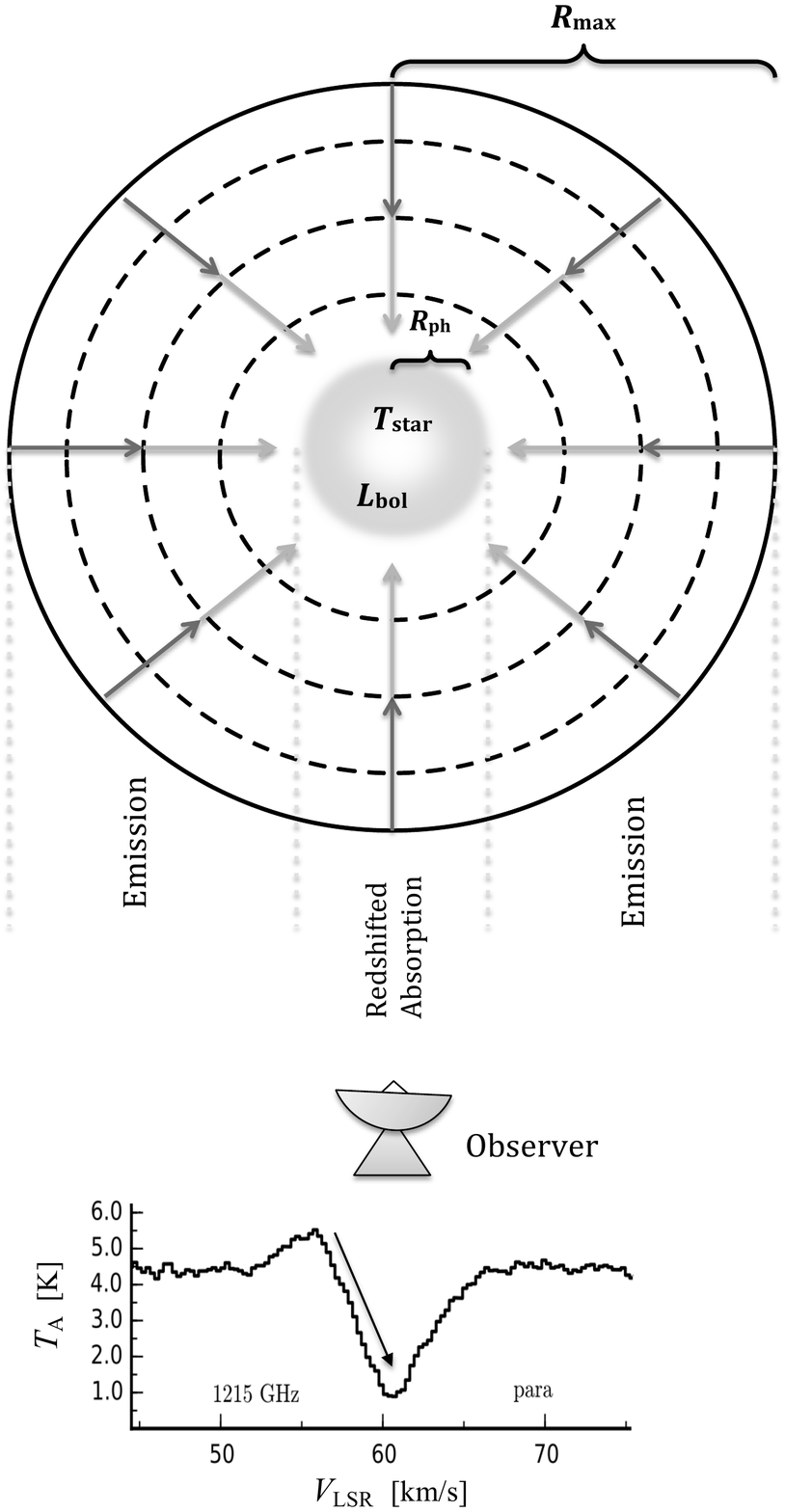}  
}
\caption{Sketch illustrating  the spherical ALI model  with shells, an inner source, 
and two velocity fields in the contracting envelope. The absorption feature is due 
to the redshifted part of the envelope in front of the strong continuum from the 
\ion{H}{II} region and the hot dust throughout the gas cloud. 
}
\label{Fig:ALImodel}
\end{figure}

The number of shell-like cells and angles used in the ray-tracing can be arbitrarily chosen. In the present work, typically 40 shells and 32 angles are used. 
Fig.~\ref{Fig:ALImodel} shows an illustration of the model and how the collapsing cloud produces an inverse P-Cygni line profile.
The ALI code produces an adjacent continuum for each shell, throughout the cloud. The centrally peaked hot dust distributions and the dust in the extended envelope emit strong infrared continuum radiation, which has a great influence on both line shapes and intensities.
The computed intensities are convolved with appropriate \textnormal{antenna beam} response functions for direct comparison with observed line profiles, corrected for beam efficiencies.

%

%
Our observed NH$_3$ line profiles show components with line widths of a few km$\,$s$^{-1}$ (cf. Table 3). The hyperfine splitting (hfs) is about 3 MHz, which corresponds to 1.5 km$\,$s$^{-1}$ for the $1_0-0_0$ line. We have therefore used spectral line data, which include the hfs splitting to generate model spectra of overlapping hfs transitions. However, collisional cross sections between hfs levels are not available for NH$_3$. 
For collisional rates, we take hfs splitting into account approximately by adopting the method outlined in \cite{Alexander} where cross sections between hfs levels can be related to purely rotational cross sections along a rotational ladder. 
As basic cross sections we here used those of \cite{danby}, which include levels $J \leq 6$ (those of \citealt{maret} only include $J \leq 3$).
\subsection*{Radiation} 
The radiation field consists of the infrared continuum radiation 
from the dust, the central source,  and the cosmic microwave background. 
The central source represents the dust photosphere 
separating  the inner region,  where diffusion determines the 
temperature gradient, from the outer region,  where balance between heating and 
cooling determines the dust temperature. 
The input parameters specifying the dust continuum are the dust temperature 
($T_{\rm dust}$), the emissivity parameter ($\kappa_{250\mu m}$), the frequency dependence of the dust emissivity ($\beta$), and the gas-to-dust mass ratio. The infrared
heating from the central source is defined by the dust photosphere's effective 
temperature ($T_{\rm star}$) and luminosity. %

The spectral energy distribution (SED)  and the physical structure of G34 were  modeled by  \cite{vdT2013} based on sub-millimeter continuum maps.   We adopt their resulting 
bolometric luminosity and source size to estimate the infrared continuum radiation from the central source. The gas temperature is to a first approximation set to be equal to the dust temperature, which holds very well for the inner part with its high densities, and lies within the uncertainties for the outer part. The power law exponent, $q$, in $T=T_0\, r^{-q}$~K for the radial temperature profile is adopted from \textnormal{a model made by \cite{garay} for molecular gas that ranges from a few hundredths of parsecs to a few parsecs, based on NH$_3$ observations at high resolution ($3\arcsec$, eight times smaller than the smallest \emph{Herschel} beam illustrated in Fig.~\ref{Fig:beamsizes})}. It is then varied together with the temperature in the outermost shell at $R_{\rm max}$, $T_0$, and the source size in order to reproduce the observed continuum within 20\%, but also the absorptions in the data. 

\subsection*{Density profile}\label{section:density}
Massive hot cores are embedded in cooler, lower density and more extended structures that are likely to be highly fragmented into filaments and dense clumps. 
Therefore the physical conditions derived assuming smooth density profiles should be taken only as representative of the bulk conditions of the clumpy emitting region.

Two cases of radial density profile are modeled. The first contains two zones, where the inner zone has constant density. The outer zone has a power-law dependence on radial distance $n=n_0\, r^{-p}$~cm$^{-3}$, in which $n_0$ (the density at $R_{\rm max}$) and $p$ are free parameters. The range of densities explored thus encompasses the various densities inferred in previous work by \cite{vdT2013}, \cite{garay},  \cite{Matthews87} and \cite{Heaton85} and includes the possibilities of having a tenuous foreground cloud or a dense extended envelope or halo. Both the boundary radius and the constant density are used as free parameters. The density profile for the second case assumes $n$ to increase continuously all the way towards the center.

\subsection*{Abundance structure}
At low gas-phase temperatures, molecules are expected to stick to the surfaces of dust grains upon collision, building up mantles dominated by water ice. 
Closer to MYSOs where the temperature is higher these ices start to evaporate. 
Since most NH$_{3}$ molecules are trapped in the water ice matrix, the bulk of ammonia will be released into the gas-phase at temperatures above 100 K, when water itself evaporates.
In the cold outer envelope, where $T<T_{\rm evap, \, \rm NH_{3}}$, the \element[][][][3]{NH} abundance 
profile is governed by reactions in the gas phase. As a result, the \element[][][][3]{NH} abundance is likely to vary in a complicated way. \textnormal{Therefore, the fractional abundance of ammonia, $X$(NH$_{3}$)~=~$n$(NH$_{3}$)/$n$(H$_{2}$), is set to be a free parameter in a step profile
(details in Sect.~\ref{Sect:Grid}).}

 \begin{figure}
 \centering
\resizebox{\hsize}{!}{ 
\includegraphics{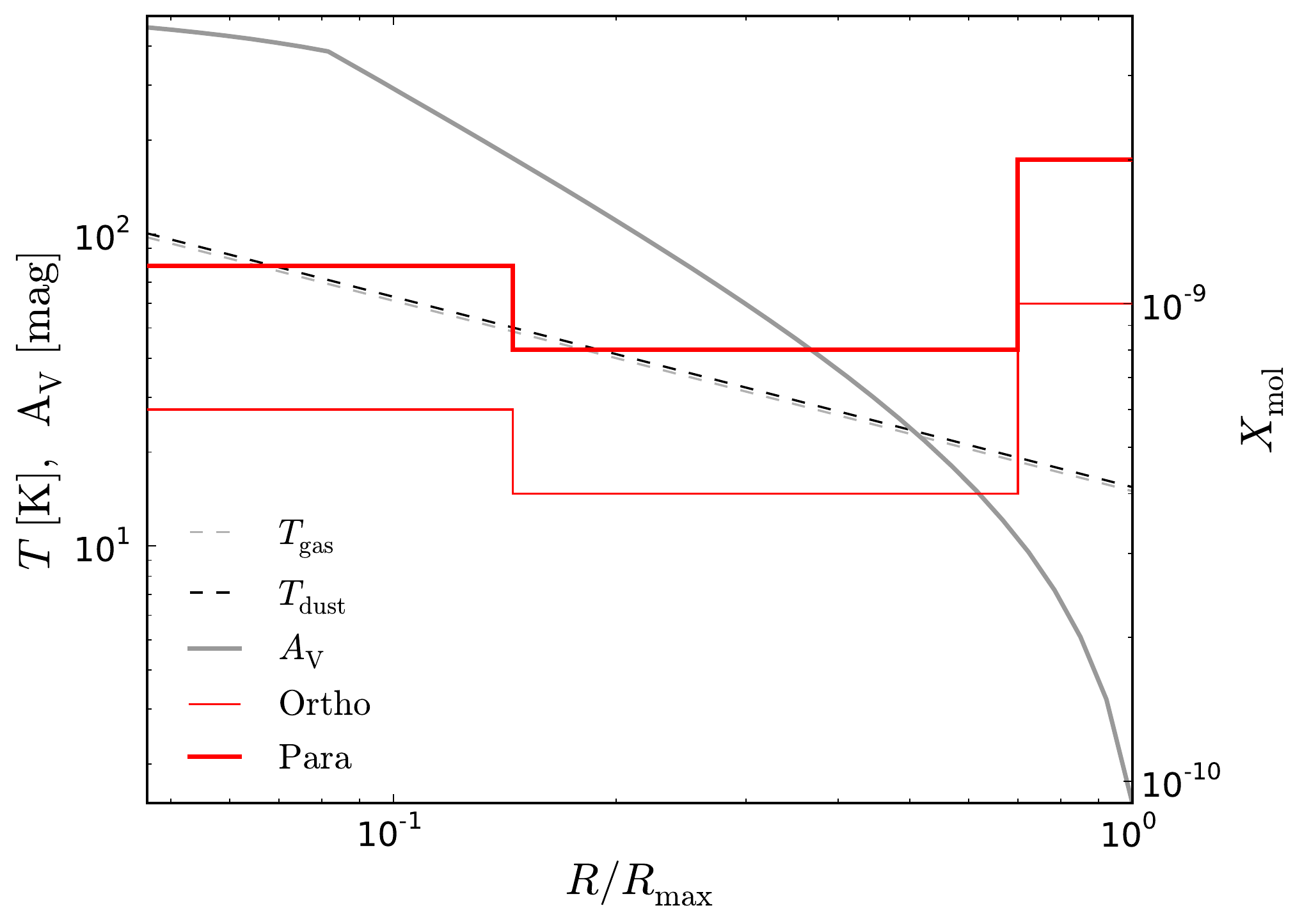}  
}
\caption{Models best-fit profile for the abundance of ortho and para NH$_3$ and the temperature, are displayed as a function of the normalized radius of the cloud. The visual extinction, $A_{\rm V}$, is proportional to the radial integral of the total hydrogen density and is displayed here as increasing from the outer boundary inwards.
}
\label{Fig:Xmol}
\end{figure} 

\subsection*{Velocity structure}\label{section:velocity}
For comparison, four different infalling velocity fields are explored where the velocity of the center is taken to be 58.1~km$\,$s$^{-1}$ : 
\emph{(i)} 
two velocity regimes, each moving with a constant velocity corresponding to the center velocities of the two absorption components distinguished in Fig.~\ref{Fig:non-uniform}, both allowed to vary over $\pm$1~km$\,$s$^{-1}$ in steps of 0.1~km$\,$s$^{-1}$ around the mean $V_{\rm LSR}$ of the Gaussian fits in Table~\ref{Table:Gaussian}  (60.6 and 63.0~km$\,$s$^{-1}$). The radius at which the two velocity fields border is described in Sect.~\ref{Sect:Grid}; 
\emph{(ii)}  
one velocity regime in free-fall similar to the modeling by \cite{wyrowski2012} and \cite{Rolffs} who used a free-fall-fraction of $f=0.3\pm0.1$.
The free-fall velocity is related to the mass interior to radius $r$, $M_{\rm in}$,  the radius at which infall begins, $a$, and scaled by the parameter $f$: 
\begin{equation}
\varv = - f\times \sqrt{2GM_{\rm in}\bigg(\frac{1}{r}-\frac{1}{a}\bigg)},
\label{eq:free-fall}
\end{equation}
where $G$ is the gravitational constant. The mass $M_{\rm in}$ includes both the mass of the central MYSOs, $M_0$, which is a free parameter (see Sect.~\ref{Sect:Grid}), and the interior mass of the collapsing cloud;
\emph{(iii)} two velocity regimes in free-fall, where $f=0.6\pm0.2$  for the inner velocity field and $f=0.3\pm0.2$ in the outer.  The radius at which the two velocity fields border and the radius at which the collapse starts are described in Sect.~\ref{Sect:Grid} ; 
and, \emph{(iv)} three velocity regimes, where the inner two velocity components are constant and described in \textnormal{(i) above}, and the outer component is in free-fall, where $f=0.3\pm0.2$.  
The radius at which the velocity fields border and start to collapse are described in Sect.~\ref{Sect:Grid}. 
No attempt has been made to incorporate a broad, redshifted component to reproduce the broad, weak emission feature in the 572 GHz line.

\subsection{Grid of models}\label{Sect:Grid}
We hold the following parameters fixed and constant through the modeled cloud: $d_{\rm source}$, $V_{\rm LSR}$, $L_{\rm bol}$, the gas-to-dust ratio, $\kappa_{250\mu m}$ and $\beta$ as given in Table~\ref{Table:model}.
\begin{table}[\!b] 
\centering
\caption{\textnormal{Parameter values} or profile of the best-fit ALI model and results.} 
\begin{tabular} {lll} 
 \hline\hline
 \noalign{\smallskip}
\textit{Parameters held constant} 			&				 &\\
 \noalign{\smallskip}	\noalign{\smallskip}
Distance to source\tablefootmark{$a$}, $d_{\rm source}$	&	3.3 	 & [kpc]\\
\noalign{\smallskip}
$V_{\rm LSR}$\tablefootmark{$b$}			&	58.1 			 & [km$\,$s$^{-1}$] \\
\noalign{\smallskip}
Envelope size, $R_{\rm max}$ 				&$1.8\times10^{18}$  &[cm]\\   
\noalign{\smallskip}
IR-continuum size, $R_{\rm ph}$ 			& $7.1\times10^{16}$ &[cm] \\
\noalign{\smallskip}
Luminosity of source\tablefootmark{$a$}, $L_{\rm bol}$	&$1.9\times10^5$ & [L$_{\odot}$]\\
\noalign{\smallskip}
IR-temperature, $T_{\rm star}$					& 120 &[K] \\
\noalign{\smallskip}
Gas-to-dust ratio, $M_{\rm gas}/M_{\rm dust}$	&	100   &\\   
\noalign{\smallskip}
Emissivity parameter, $\kappa_{250\mu m}$		&	20 &[cm$^2$ g$^{-1}$]   \\   
\noalign{\smallskip}
Dust frequency dependence, $\beta$				&	1.8  & \\   
\noalign{\smallskip} 
Ortho to para ratio, OPR  &  0.5\\ \noalign{\smallskip}
\\
\textit{Parameters profile} \tablefootmark{$c$}	& 	&\\
\noalign{\smallskip}  \noalign{\smallskip}
Gas temperature, $T_{\rm g}$ 					&	$15.0\,r^{-0.61}$&[K]   \\   
\noalign{\smallskip}  
Dust Temperature, $T_{\rm d}$ 				&	$1.03\times T_{\rm g}$&[K]   \\   
\noalign{\smallskip}
H$_{2}$ density, $n(\rm H_{2})$			&	\\
$r<0.075$								& 	$2.3\times10^6$ 		 &[cm$^{-3}$] \\
$r>0.075$ 							&	$1.8\times10^4\,{r}^{-2.3}$ &[cm$^{-3}$] \\
\noalign{\smallskip}   
Velocity profile, $\varv_{\rm infall}$			&	\\
$r< 0.2 $							&	 5.3 &[km$\,$s$^{-1}$]\\
$0.2 < r > 0.7$						&	 2.7 & [km$\,$s$^{-1}$]\\
$r> 0.7 $							&	$\left\{\begin{tabular}{l} $f$ = 0.3 \\  $M_0 = 1$  \\      $a  = 1$   \end{tabular}\right.$ & [M$_{\odot}$] \\ 
\noalign{\smallskip}
Micro turbulence, $\varv_{\rm turb}$ 			& 	 \\   
$r< 0.9 $						& 	1.0& [km$\,$s$^{-1}$] \\ 
$r> 0.9 $								& 	1.6& [km$\,$s$^{-1}$] \\ 
\noalign{\smallskip}
Abundance, $X(\mathrm{NH}_3)$	 &	   \\ 
$r<0.145$	 & $1.2\times10^{-9}$&   \\  
$0.145<r<0.7$	 & $0.8\times10^{-9}$&   \\ 
$r>0.7$ 	&  $2.0\times10^{-9}$\\	
\noalign{\smallskip}  
\noalign{\smallskip} \noalign{\smallskip}
\hline \noalign{\smallskip} \noalign{\smallskip}
\textit{Results} 	& 	&\\
\noalign{\smallskip} 
\noalign{\smallskip}
$N(\rm p$-$\mathrm{NH}_3)$	 &	$9.1\times10^{14}$&[cm$^{-2}$]  \\  
$N(\rm o$-$\mathrm{NH}_3)$	 &	$4.6\times10^{14}$&[cm$^{-2}$]  \\  
\noalign{\smallskip} 

$N(\mathrm{H}_2)$	 &	$8.4\times10^{23}$&[cm$^{-2}$]  \\ \noalign{\smallskip} 
\noalign{\smallskip}

$M$(G34)\tablefootmark{$d$} 	 &	2\,700&[M$_\odot$]  \\ \noalign{\smallskip}
\noalign{\smallskip} 

$\dot{M}_\mathrm{acc}$\tablefootmark{$e$}   & \\ 
$r= R_{ph}$		 &  $4.1\times10^{-3}$  & [M$_{\odot}$~yr$^{-1}$]  \\
$r= 0.075R_{max}$	 &  $4.5\times10^{-2}$  & [M$_{\odot}$~yr$^{-1}$]  \\
$r= 0.2R_{max}$	 &  $1.7\times10^{-2}$  & [M$_{\odot}$~yr$^{-1}$]  \\
$r = 0.7R_{max}$  	&   $4.5\times 10^{-3}$   & [M$_{\odot}$~yr$^{-1}$]\\	
\noalign{\smallskip}  
\noalign{\smallskip} \noalign{\smallskip}
\hline 

\label{Table:model}
\end{tabular}
\tablefoot{
\tablefoottext{$a$}{\cite{vdT2013}}
\tablefoottext{$b$}{\cite{wyrowski2012}}
\tablefoottext{$c$}{The parameters vary with \textnormal{normalized} radius, $r = R/R_{\rm max}$.}
\tablefoottext{$d$}{Mass inside $R_{\rm max}$.}
\tablefoottext{$e$}{See equation~(\ref{massacc}).}
}
\end{table} 
%
We vary the envelope size, $R_{\rm max}$, and the effective IR-temperature of the star, $T_{\rm star}$, to find the best combination before we hold these values constant. 
\textnormal{For a fixed luminosity, the size and stellar temperature can vary: the higher the value of $T_{\rm star}$, the smaller the value of $R_{\rm ph}$, and vice versa. Values of  $T_{\rm star}$ between 70 and 1000~K were tested.  The range of maximum envelope sizes was $1\times10^{18}$ to $3\times10^{18}$~cm. This range includes the sizes estimated by \cite{vdT2013} and \cite{Rolffs} as well as the radius derived from the comparison of Odin and Herschel measurements at 572~GHz ($\S$\ref{sect:size}). The continuum radiation from mm to infrared wavelengths stems from heated dust grains.}

The following parameters have a profile structure through the cloud that \textnormal{is} varied:
\begin{itemize}
\item Temperature: $T_0$, in $T\,$=$\,T_0\,{r}^{-q}$, is varied between 5 and 30~K, in steps of 2~K, and $q$, is varied $0.6\,\pm$0.2 in steps of 0.01. 
\item Density: $(i)$ \emph{continuous profile:} \textnormal{ $p$, in $n\,$=$\,n_0\,{r}^{-p}$}, is varied between 1.5 and 2.5 in steps of 0.1 along with $n_0$,  \textnormal{which is allowed to vary between $10^3$ and $10^5$ cm$^{-3}$}. $(ii)$ \emph{Inner profile:} the boundary radius and the density are varied, between 0.05$R_{\rm max}$ and 0.1$R_{\rm max}$ and between $10^5$ and $10^8$ cm$^{-3}$, respectively. \emph{Outer profile:} is the same as the \emph{continuous profile}.
\item Abundance: (i) Values between $10^{-10}$ and $10^{-7}$ are varied as a free parameter in a constant profile through the whole cloud. (ii) Same range of abundance is used to vary for a two and three-step profile, where the radius at each step is varied as a free parameter. The OPR for any profile is varied between 0.3 and 1.0.
\item Velocity: \emph{(i)} For the two and three velocity regimes profile the intersecting radius is varied between 0.04$R_{\rm max}$ and 0.9$R_{\rm max}$, with steps of 0.01$R_{\rm max}$. \emph{(ii)} For a free-fall collapse the mass of the UC \ion{H}{II} region is varied between 1 and 2000~M$_{\odot}$. The collapse is assumed to be initiated at $a= R_{\rm max}$. 
\item Micro turbulence: $(i)$ A constant $\varv_{\rm turb}$ profile is set through the whole cloud and varied between 0.5 and 2~km$\,$s$^{-1}$.
$(ii)$ A step profile with a constant inner and outer $\varv_{\rm turb}$ is varied together with the boundary radius, as free parameters. The profile in the outer envelope is set to a higher value compared to the inner, as measured by \cite{Liu, Coutens} for G34 and by \cite{Herpin, Caselli}  for other massive protostars.
\end{itemize}

\subsection{Uncertainties of the parameters}\label{sect:chi2}
We use the chi-square ($\chi^{2}$) method to get a statistical comparison of the qualities of fits. For each line and model we compute
\begin{equation}
\chi^{2}_{\rm line} = \sum^{n_{\rm ch}}_{i = 1}\frac{(o_i-m_i)^{2}}{\sigma^{2}}
\end{equation}
where $n_{ch}$ is the number of channels between \textnormal{$V_{\mathrm{LSR}}= 54$ and 68~km$\,$s$^{-1}$} where  the absorption is located for all lines and models. $o_{\rm i}$ and $m_{\rm i}$ are the observed and modeled continuum subtracted intensities, respectively. 
The observational error is dominated by calibration uncertainties (assumed to be 14~\%, thus representing the band with the highest quadratic combination of noise presented in Sect.~\ref{section:obsNred}) rather than by thermal noise that is $<$10~\%. Therefore $\sigma$ represents the calibration uncertainty multiplied with the maximum intensity, $T_{\rm mb}(\rm max)$, in the spectra (with continuum, $T_{\rm C,\rm mb}$). This method is adopted from \cite{Rolffs}. 
To compare the models we sum $\chi^{2}_{\rm line}$ over all transitions, except for the 572~GHz transition \textnormal{since we have limited possibilities to model the broad redshifted emission with ALI.} The model with the \textnormal{lowest} value of $\sum{\chi^{2}_{\rm line}}$ 
determines the best-fit model. 
A model is assumed to successfully reproduce the observed line profile if the continuum intensity of the best-fit model is within 20\% of  the observational value. 

\section{The best-fit model}
\label{Section:Result}

\begin{figure}[\!htb] 
\resizebox{\hsize}{!}{ 
\includegraphics{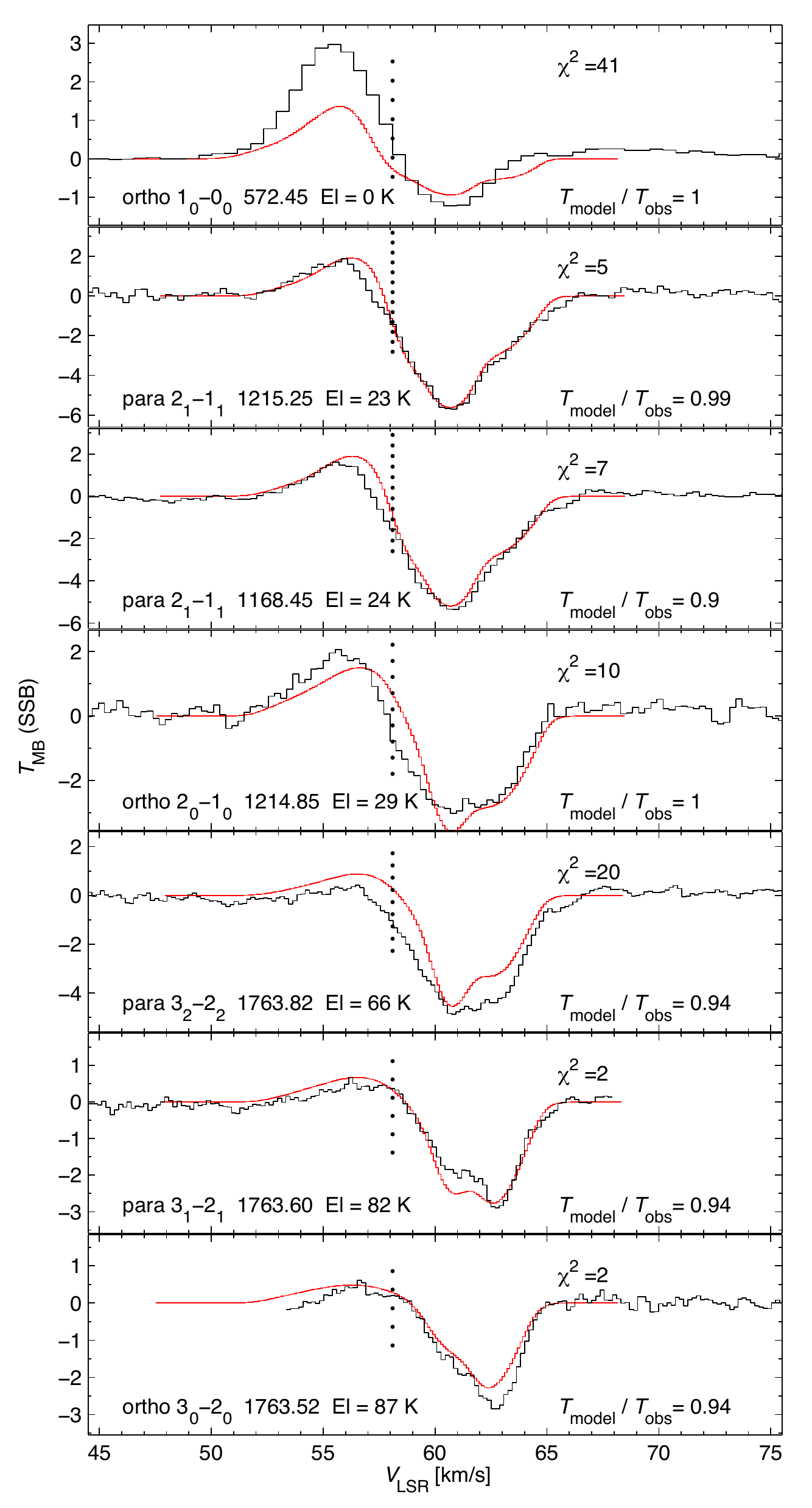}
}
\caption{
\emph{Three velocity regimes with two inner constant profiles and an outer free-fall profile.} The inner and outer constant velocity profiles have a velocity of 5.3 and 2.7~km$\,$s$^{-1}$, respectively, and the free-fall profile uses $f$ = 0.3, $\mbox{$M_0 = 1$~M$_{\odot}$}$ and $\mbox{$a$ = 1}$. 
Observed \textit{Herschel}-HIFI spectra in black compared to the best-fitting model line profiles in red. 
The ratio of the modeled and observed continuum at the frequency of each line is given in respective legend. 
The vertical black dotted line marks the systemic velocity of G34. $\sum{\chi^{2}_{\rm line}}$ = 46
, over $\Delta V_{\mathrm{LSR}}= 54-68$~km$\,$s$^{-1}$.
}
\label{Fig: model}
\end{figure}

\textnormal{
The physical parameters of the best-fitting model are presented in Table~\ref{Table:model}. The line profiles and continuum fluxes of this model are compared with the observed spectra in Fig.~\ref{Fig: model}. The robustness of the model is discussed in Sect.~\ref{sect:param}.
The models used to describe G34 by \cite{wyrowski2012} and \cite{Coutens} are discussed and compared with the best-fit model in Sect.~\ref{section:comparison}. }

Six of the absorption line profiles in Fig.~\ref{Fig: model} are very well reproduced between $V_{\mathrm{LSR}}$ = 60 and 66~km$\,$s$^{-1}$. \textnormal{Two of the emission line profiles are, however, somewhat overestimated, but one could expect a perfect match only if the observed region were more uniform and unstructured within the projected beam. 
How well the model is able to reproduce the observed continuum level is shown in the lower right corner of each spectrum, for the best-fit model always within 10\%.
It is a challenge to model the fundamental ortho 572~GHz line simultaneously with the higher excited transitions. Both too little emission and too little absorption are found, which is speculated to be due to the one-dimensional spherical symmetry of the model but also our assumed simplifications of the source structure, as further explained in Sect.~\ref{sect:limitations}. 
 }

\textnormal{There is evidence of gradients in the density, gas and dust temperature, micro turbulence and ammonia abundance, within the inflowing molecular gas associated with G34. The scale of the gradient from observational data, is between 0.2~pc (with the best resolution being 12$\arcsec$ at 3.3~kpc) and  0.6~pc (36$\arcsec$ at 3.3~kpc). }
For the best-fit model it stretches longer, between 0.023~pc and 0.583~parsec.  
\textnormal{\cite{garay} proposed a physical structure of ammonia, based on interferometric observations at cm wavelengths, which shows that the best-fit model probes a region between a \textit{Hot ultra-compact core} and a \textit{Molecular core}.}

%
%

\subsection{Parameters}\label{sect:param}
\subsection*{Temperature profile}
\begin{figure}[\!htb]
\resizebox{\hsize}{!}{ 
\includegraphics{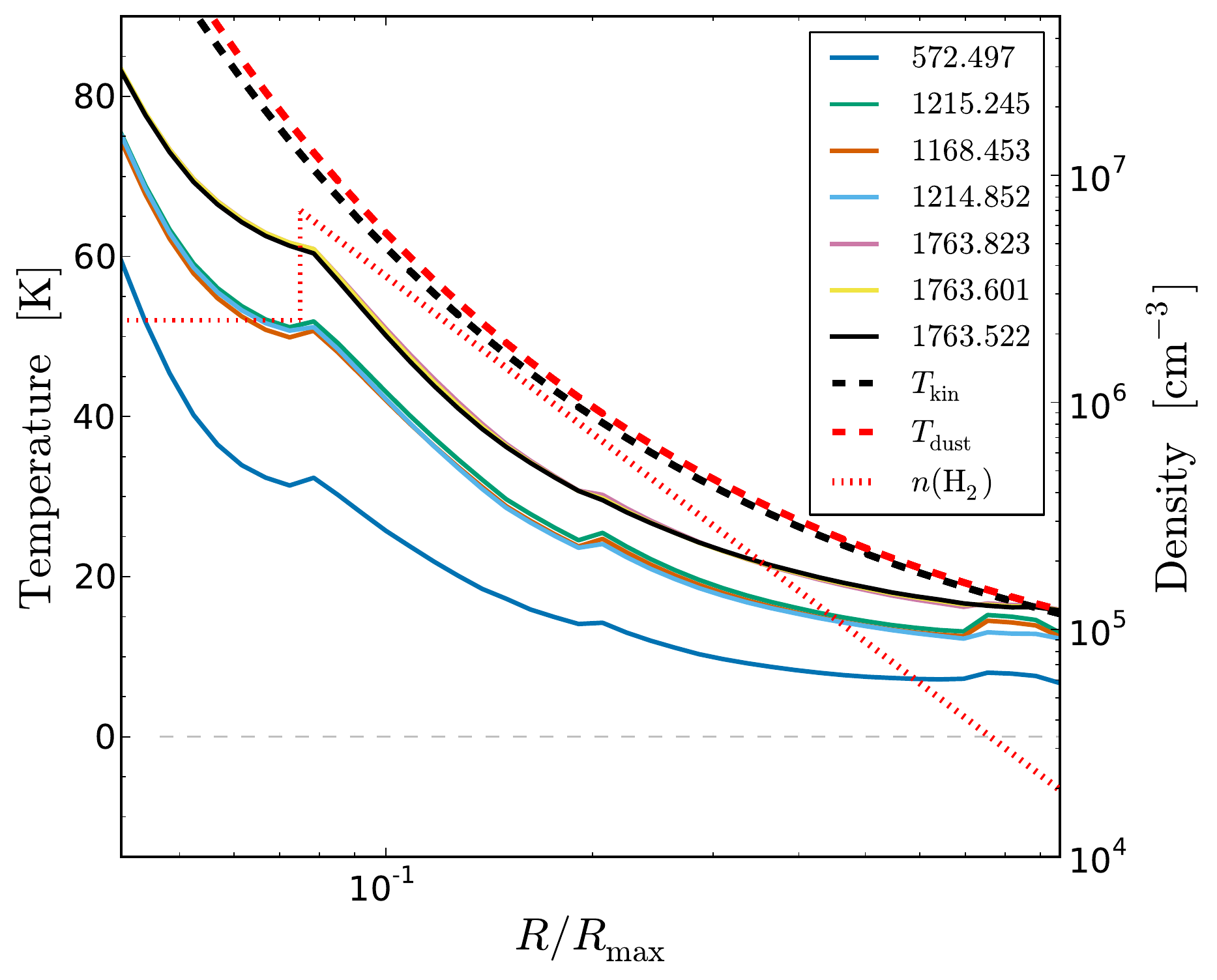}  
}
\caption{Excitation temperatures in the best-fitting model as a function of the normalized radius. The gas temperature, $T_{\rm g}$, dust temperature, $T_{\rm d}$, and density, $n$(H$_2$), are also plotted for comparison.
We use the same color code as in Fig.~\ref{Fig:Energy}.}
\label{Fig:Tex}
\end{figure}

The modeled $T_\mathrm{ex}$, $T_\mathrm{K}$ and $n(\mathrm{H_2})$ 
are plotted as functions of normalized radius in Fig.~\ref{Fig:Tex}. 
\textnormal{The difference between excitation temperature and kinetic temperature is the result of departures from LTE throughout the envelope. This shows that the density is too low for collisional excitation to dominate over radiative processes. }
%

\subsection*{Density profile}
\textnormal{
The combined emission and absorption profiles are well reproduced by a step-wise density profile, which permits two distinct infall velocities to be included. The best-fitting model has a central flattening of the density distribution joined to a power-law profile that decreases outwards\footnote{similar to what \cite{Heaton93} derived using HCO$^{+}$}. Adoption of a continuous density profile throughout the envelope results in a somewhat poorer fit ($\sum{\chi^{2}_{\rm line}}$ increases from 46 to 50), with excess absorption in the 1214 and 1763.6~GHz lines and excess continuum near all lines. An example of a continuous model is shown in Fig.~\ref{fig:constdens}. 
The central flattening in our best-fit density profile, illustrated in Fig.\ref{Fig:Tex}, is located at $<$0.04~pc. 
\textnormal{An increase of the radius of this region decreases the emission, absorption and continuum } for all the lines (especially for the lines of highest excitation) and vice versa when decreasing its radius. }

The flat and lower density profile for the inner envelope, of $\lesssim 9000$~AU, can be explained by: 
\begin{enumerate}
\item A disc of molecular gas that is formed from the collapse of a massive, rotating fragment embedded in an extended molecular cloud as suggested by \cite{garay} and \cite{Garay86}. The size of the possible disc ($\lesssim 9000$AU) agrees well with estimated sizes of disks associated with young massive objects (single protostars or clusters) \textnormal{showing} signs of outflows but no jets \citep{Garay1999}. 

\item \textnormal{The relation between the inner envelope} and the contraction process of gas and dust, causing the disk to spin-up to high velocity gradients, which \cite{Keto} observed G34 and W3(OH) (an UC \ion{H}{II} region) to have by studying rotational motions of the emitting gas (NH$_3$(1,1) transition). A similar result was observed toward at least five other UC \ion{H}{II} regions by \cite{klaassen2009, Keto-Haschick}.

\item A cloud-cloud collision in our line of sight between two components with two different density profiles, moving inwards. 
Alternatively one cloud moving inwards and the outer outwards, which \textnormal{would be} in agreement with what \cite{Coutens} found in the inner region from HDO observations and modeling.

\item \textnormal{Part of the ammonia being dissociated closest to the central \ion{H}{II} region (see Fig.~\ref{Fig:beamsizes} for scale comparison), thus not tracing the total gas mass.}

\end{enumerate}

The integrated H$_2$ mass within the spherical shells \textnormal{of our best-fit model} (see Table~\ref{Table:model}) is $\sim$50\% higher than the 1800~$M_{\odot}$ found by \cite{vdT2013} in \textnormal{their modeling of low-excitation water lines}. 

\subsection*{Abundance structure \&  Column density}
The best-fit ammonia abundance profile, adopted in our model, is shown in Fig.~\ref{Fig:Xmol} together with the visual extinction and temperature \textnormal{profiles}. 
\textnormal{
It suggests an ammonia abundance of $\mbox{$1.2\times10^{-9}$}$ and $\mbox{$2.0\times10^{-9}$}$ in the inner and outer regions, respectively, but a lower abundance half way into the envelope. }

Since the temperature in the modeled region does not reach 100~K even in the innermost parts, a major fraction of the ammonia is expected to be bound in the ices throughout the envelope. 
However, external UV-photons may penetrate the envelope and photodesorb ammonia molecules from dust grains in the outer parts where the extinction is low, which may explain the increase in ammonia abundance we see for $A_{V}<15$~mag. In addition, desorption by cosmic ray-induced UV-photons may be responsible for some of the observed gas-phase abundance throughout the observed envelope.

\textnormal{The abundance in the outer region of our best-fit model is similar to what is measured in translucent gas \citep{Persson12}, where the temperature and density are similar to the conditions modeled for the outer envelope of G34. 
}

\textnormal{Comparing our low derived abundances} with the chemical models for a relatively dense gas ($n_{\rm H_2}$=$2\times10^{5}$~cm$^{-3}$, $T$ = 16~K) presented in \cite{Coutens}  suggests that the molecular gas in G34 is not  in steady-state, with an age of \textnormal{only about} $10^5$~years. This can be the result of the short evolutionary timescales of massive stars. 
Alternatively, \textnormal{the low derived abundances} might be due to the assumption of equal gas and dust temperatures, which may break down in the low-density region.

\textnormal{The column density derived (in Sect.~\ref{Sect: column}) to be the lower limit is approximately twice as high as that found by \cite{wyrowski2012}, but five times lower than suggested by our ALI modeling.}

\subsection*{Ortho-to-para ratio}
We have for the first time derived an OPR below the high temperature limit of unity in a star forming region. \textnormal{The} modeled value of 0.5 is in agreement with the results of \cite{Persson12} who found an ammonia OPR of ~0.5~$-$~0.7 in translucent interstellar gas. 
Since no radiative transitions are allowed between ortho and para states an OPR below unity has been considered \textnormal{to be} impossible until recently. It can be explained by the nuclear spin selection rules in a low-temperature para-enriched H$_2$ gas since this naturally drives the OPR of nitrogen hydrides to values below their respective statistical high temperature limit \citep{OPR, LeGal}.
Recent estimates of OPR ratios from chemical models by \cite{LeGal}  are \textnormal{lower} than unity ($\sim$~0.4~$-$~0.7), in good agreement with our \textnormal{modeling} results.

\subsection*{Velocity structure}\label{result:velocity}
\textnormal{The} best-fit model has both turbulence and infall that vary with radius (see Fig.~\ref{fig:velprofile}). Analysis of the absorbing material towards G34 suggests a scenario where \emph{two} and possibly a third gas envelope are moving inwards toward the central region, where the two inner ones have  constant velocities and the third \textnormal{outermost envelope (where r~>~0.7~R$_{\rm max}$)} has a free-fall velocity profile. 

In order to produce an acceptable fit at least two velocity regimes are needed. Either two constant velocity regimes, with infall at 5.3~km$\,$s$^{-1}$ for r~<~0.2 and 2.7~km$\,$s$^{-1}$ for r~>~0.2, shown in Fig.~\ref{Fig: 2constvel}, or two in free-fall, shown in Fig.~\ref{fig:fff0.6fff03M400} (see velocity profile shown in Fig.~\ref{fig:velprofile}). The former models the absorption profiles well, especially at higher velocities. The latter models the emission and absorption at lower velocities better, and \textnormal{require free-fall velocity} fractions of $f$~=~0.7 and 0.4 for the inner and outer velocity field respectively. In addition, it needs the mass contained in the compact central object (within radius $R_{\rm ph}$) to be on the order of $\sim$400~$M_{\odot}$. 

\textnormal{Thus, a combination of these two models produces the best-fit as shown in Fig.~\ref{Fig: model}. If the mass of the central source, $M_0$, is taken to be less than 300~M$_{\odot}$ it will not affect the free-fall velocity ($\varv_{\rm infall}$) profile in the outer envelope, and a higher mass will shift the absorption minima to higher velocities and increase the sum of $\chi^{2}_{\rm line}$. Therefore, $M_0$ is set to be equal to 1~M$_{\odot}$}.

\subsection*{Micro turbulence}
\textnormal{The "two-step profile" for the turbulence (see Table~\ref{Table:model} for details) provided better fits for all the line profiles, compared with setting the micro turbulence as constant through the source. Thus we adopted an increase of the micro turbulence from unity to 1.6~km$\,$s$^{-1}$ in the outer envelope \citep[cf.][]{Liu,Herpin,Caselli}. }
\textnormal{This parameter adjusts the detailed profile shapes.}

\subsection{Modeling}\label{section:comparison}
\textnormal{The first step of modeling the seven ammonia transitions in ALI was to adopt the parameters used by \cite{wyrowski2012} to model the infalling component at about 61~km$\,$s$^{-1}$ observed in the $3_{2,-} - 2_{2,-}$ NH$_3$ transition at 1810~GHz (see Fig.\ref{Fig:Energy}), obtained by SOFIA.
Its critical density is similar to those of the 1764~GHz transitions and therefore might share the same environmental origin.}
\textnormal{
We implemented the same temperature and density profile as given in \cite{Rolffs} and a free-fall velocity field as given in \cite{wyrowski2012}, where $\mbox{fff = 0.3, $M_0 = 20$~M$_{\odot}$}$  and $\mbox{$a$ = 1000}$. While the 61~km$\,$s$^{-1}$ absorption of the 1764~GHz line could be reasonably fit, the model overestimates the blue-shifted emission and cannot reproduce the red-shifted absorption component at 63~km$\,$s$^{-1}$. This led us to the changes in the physical structure discussed in Sect.~\ref{result:velocity} which resulted in our best-fit model.
} 

\textnormal{
Using the best-fit density profile (see Sect.~\ref{result:velocity}) when adopting the one component free-fall velocity, the infalling component at about 61~km$\,$s$^{-1}$ is poorly modeled. See velocity profile,  $\varv_{\rm 1freefall}$, shown in Fig.~\ref{fig:velprofile} and modeled line profiles shown in Fig.~\ref{fig:fff0.3M20}. }

\textnormal{The best-fit model also reproduces well the profile of the 1810 GHz transition (Fig.\ref{Fig:bestfit1810}). In fact, the fit is somewhat better than for the corresponding \emph{Herschel}-HIFI transition at 1763.8 GHz (i.e., $\sum{\chi^{2}_{\rm line}}$ = 13 vs 20). The 30\% discrepancy in continuum flux could well be within the calibration uncertainty of that observation.
}

\textnormal{The infalling component at about 61~km$\,$s$^{-1}$ was modeled by \cite{wyrowski2012} for NH$_{3}$ but also by \cite{Coutens} for HDO. 
At a corresponding radius of r~=~0.066~$R_{\rm max}$ in the best-fit model, \cite{Coutens} modeled an expanding region set to 4~km$\,$s$^{-1}$.
When this velocity is applied to the best-fit model at r~<~0.075, the absorption components are not affected, but the peak of the emissions is narrower. Still, it does not produce a lower $\sum{\chi^{2}_{\rm line}}$ value than the best-fit model. As a result, we can neither confirm nor refute the presence of an expansion in the inner parts of the envelope. }
%

\section{The mass accretion process}
\label{section:discussion}

The very first stage of the formation of massive OB stars is believed to be initiated by contraction and fragmentation of dense molecular cores in approximate hydrostatic equilibrium \citep{Garay1999}. 
The question is then how a molecular core evolves to produce one or several massive stars? 
High amounts of mass must be able to feed the star on short time-scales despite the strong radiation pressure. Models have shown that transfer of matter through a protostar-disk system is very efficient in order to accrete matter to the star (see Sect.~\ref{introduction})

Accretion can proceed if its rate is \textnormal{higher} than the mass that can be supported by radiation pressure from the luminous protostars. \textnormal{Our highly resolved multitransition study confirms the inward motion shown by Wyrowski et al. 2012, but it also shows increased velocity with increased excitation. With a spherically symmetric accreting envelope, the best-fit model can reproduce the line profiles and continuum. } \textnormal{The size of the accreting envelope is clearly larger than the separations of MYSOs in the embedded cluster. The mass in each shell of the envelope of our model can be interpreted as separate, each moving inwards. This result is consistent with the competitive accretion model \citep{bonnell1997, bonnell2001}, in which many YSO compete for the same reservoir of material.}

\textnormal{The mass accretion rate at any radius $R$ can be estimated by means of the method of \cite{Beltran}}
\begin{equation}
\dot{M}_{\rm acc} = 4\pi \,R^{2}\,m_{\rm H_2}\,n_{\rm H_2}\,\varv_{\rm infall}, 
\label{massacc}
\end{equation}
where $m_{\rm H_2}$ is the  mass of the H$_2$ molecule, $n_{\rm H_2}$ is the gas volume 
density of  molecular hydrogen, 
and $\varv_{\rm infall}$ is the infall velocity as a function of the radius $R$ as described in Table.~\ref{Table:model}. 

\begin{figure}[\!htb] 
\resizebox{\hsize}{!}{ 
\includegraphics{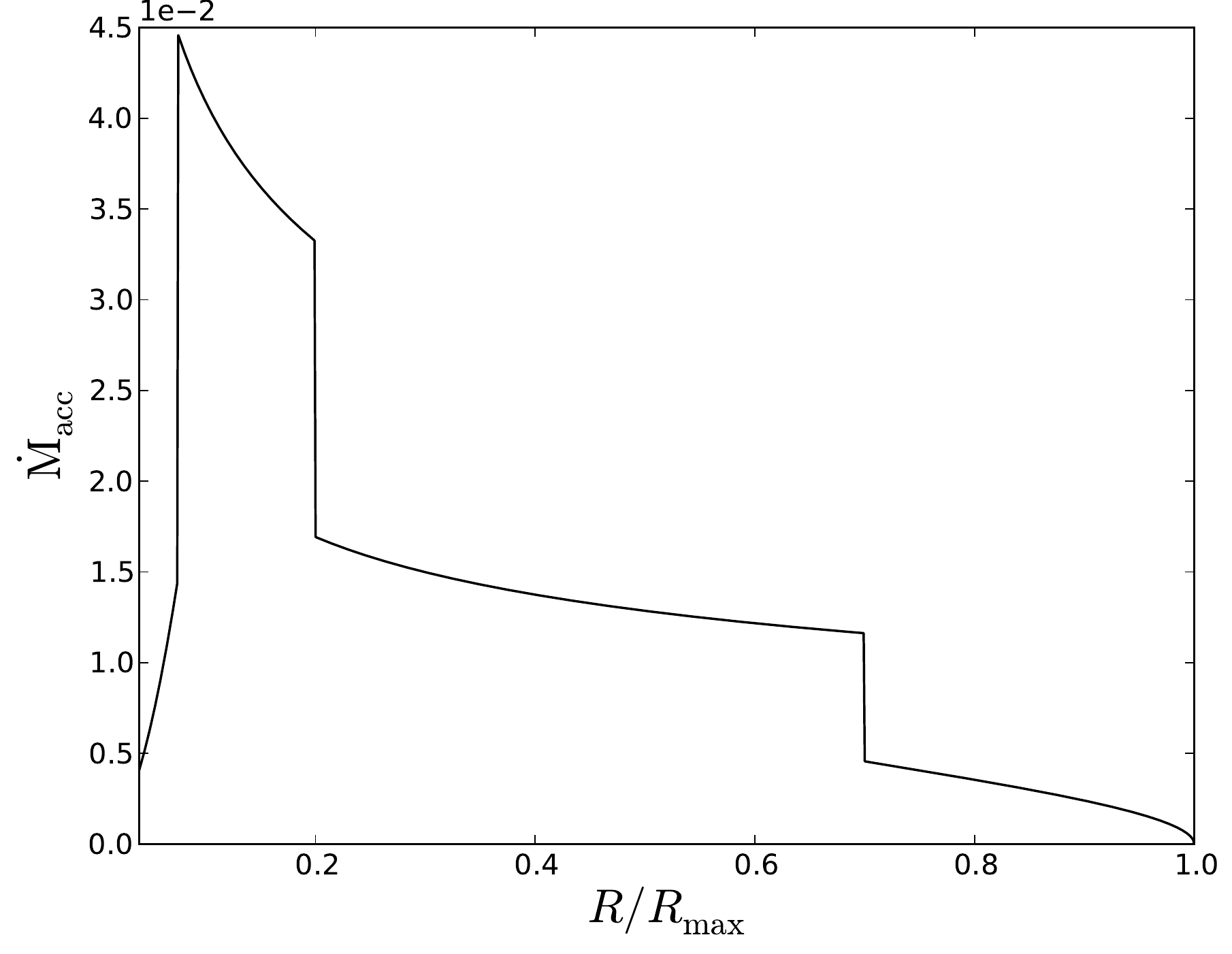}} 
\caption{Resulting mass accretion profile, $\dot{M}_{\rm acc}$ [~M$_{\odot}$yr$^{-1}$], obtained in the best-fit ALI model as a function of the normalized radius of the cloud.}
\label{fig:massaccretionprofile}
\end{figure}

\textnormal{The inferred mass accretion rate starts from zero at $R_{\rm max}$ and increases inwards together with the free infall velocity and the $n_{\rm H_2}$. }
When the velocity in the outer envelope is free-falling the mass accretion rate ($\mbox{$4.5\times10^{-3}$~M$_{\odot}$yr$^{-1}$}$) supports estimates made by \citet{wyrowski2012}, \citet{klaassen} and \citet{Liu}.
As the velocity jumps up to 2.7~km$\,$s$^{-1}$, at r = 0.7$R_{\rm max}$, the mass accretion rate also jumps up to $\mbox{$1.2\times10^{-2}$~M$_{\odot}$yr$^{-1}$}$. 
Within the inner 20\% of the envelope, where the infall velocity is 5.3~km$\,$s$^{-1}$, $\dot{M}_{\rm acc}$ increases to $\mbox{$3.0\times10^{-2}$~M$_{\odot}$yr$^{-1}$}$. At 0.075$R_{\rm max}$, where the number density and velocity are the highest, the mass accretion rate reaches its maximum and is $\mbox{$4.5\times10^{-2}$~M$_{\odot}$yr$^{-1}$}$, just before the profile of $n_{\rm H_2}$ decreases to $2.3\times10^6$~cm$^{-3}$.
\textnormal{The apparent decrease in mass-accretion rate at r~<~0.075~$R_{\rm max}$, where the gas density has been artificially truncated, signifies that the accretion must shut off at the surface of the IR-continuum object.}
The resulting profile of the mass accretion rate is shown in Fig.~\ref{fig:massaccretionprofile}.

We speculate that the modeled flat inner density profile \textnormal{may} originate from \textnormal{an} accretion disk, which is fed by the infalling envelope. This is also supported by observations made by \cite{Keto} of G34 that suggest high rotational velocities.
From the radial size of the inner constant density profile the accretion disk is about 9000~AU in radius, which could mean that the possible accretion disk is large enough to enclose several protostars. Protostars surrounded by an accretion disk, with an ensemble of molecular clumps falling inwards, agrees well with the dynamical competitive accretion model.

\textnormal{The overall high infall velocities through the envelope and the high mass infall rates for G34 are expected for more evolved MSF regions compared with the stages prior to the hot core phase \citep{Liu}. The mass accretion rates are high enough (>$10^{-3}$~M$_{\odot}$yr$^{-1}$) throughout the envelope to quench the \ion{H}{II} region in G34 \citep{Walmsley, Wolfire} and to overcome the radiation pressure from the forming (proto) cluster. }

If there existed an outward motion in the innermost region, as suggested by \cite{Coutens}, then the mass accretion rate would be zero in this region. Non-spherical symmetric motions of gas and dust would be needed to feed the protostars in the cluster, which is consistent with the competitive accretion model.

We stress that the derived high mass accretion rates assume spherically symmetric accretion, which may not be true. The true accretion rate could be different if the accretion is non-uniform or lacks spherical symmetry. 
If this is the case, the minimum fraction \textnormal{of the spherically symmetric mass accretion rate needed for the shell to be able to quench the \ion{H}{II} region at radius,  $R_{\rm ph}$, $0.075R_{\rm max}$, $0.2R_{\rm max}$ and 0.7$R_{\rm max}$, are: 24\%, 2.2\%, 3.3\%, 22\%, respectively. 
However, it is beyond the scope of this paper to consider the different distribution of radiation pressure on the accreting clumps compared to the case of spherically symmetric accretion.}

\textnormal{The central mass of the source has a great influence on the free-fall velocity field(s), thus the mass accretion rate.
From literature, the masses estimated from high-resolution observations of the central core of G34 are 76 to 360~M$_{\odot}$  \citep{Liu, watt, Garay86}. The observed core has a diameter of $2\times R_{\rm ph}$ that spans about 1.4~\arcsec on the sky at 3.3 kpc distance. 
This is in agreement of the upper limit of approximately 300~M$_{\odot}$ in the central source found in our best-fit model.}


\subsection{Modeling limitations}\label{sect:limitations}
Since our model is based on spherical symmetry it \textnormal{has limited abilities to take into account} clumpy distributions of gas and dust and asymmetric outflow. 
The compact \ion{H}{II} region in G34 is far from spherically symmetric \textnormal{since component C has a cometary tail on large scales (see Fig.~\ref{Fig:beamsizes})} where most likely the bulk of the emission and absorption of the 572~GHz line is produced. \textnormal{The shape makes the density distribution non-symmetric, having lower density to the right part of the tail compared to the left side. The influence of these} components may result in an underestimation of the absorption.  The resulting line profile is also affected by surrounding molecules, which is beyond the scope of our model.
%
%
\textnormal{The broad redshifted emission of o-NH$_{3}$ (572~GHz) and o-H$_{2}$O (557~GHz) trace a broad outflow component\footnote{known to exist in G34 \citep{vdT2013, Garay86}}, estimated to be large enough to support the scale of a molecular champagne flow as a result of the evolved \ion{H}{II} region, expanding as a bubble until it reaches less dense ISM and breaks. 
The morphology, age and driving source of the subparsec-scale outflow components \cite[see Sect.~\ref{introduction}, ][]{hiroshi,klaassen} are still unknown but have suggested to be associated with individual YSO harboring a turbulent disk \citep{Heaton85}, as observed in low-mass objects. 
They should give rise to bipolar collimated outflows, however the reason why we are only detecting the red-shifted part must be that the the opposite outflow lobe is directed tangentially to our line of sight and therefore is invisible to the telescope.}
%

The high luminosity of G34 implies that massive (proto)stars are the heating engine, although we have modeled the source with only one central heating source. Multiplicity could indeed partly explain the deviations between model and data, since the sources are clustered \textnormal{from the onset}. A more realistic picture of a forming star cluster would include multiple heating sources, \textnormal{which is impossible for our model to cover.}

Limited computational power prevented tests of all model combinations and allowed only a simple chemical structure, not computed from chemical models.
The gas-to-dust mass ratio, emissivity parameter and the dust frequency dependence are assumed to be constant throughout the source, although they are likely to vary (e.g., due to evaporation of ice mantles).

\section{Conclusions}
\label{section:conclusion}
In this paper we have presented seven velocity-resolved rotational lines of ammonia toward the ultra compact HII region G34.26+0.15.
The mixed absorption and emission in the line profiles was successfully modeled with the radiative transfer code ALI \textnormal{(see Fig.~\ref{Fig: model})}.

The narrow-emission part of all the seven ammonia line profiles and the absorption for the ground-state line, are quite well reproduced, except for the emission part \textnormal{for a few transitions.} This can be the result of assuming that the enveloping matter is spherically symmetric. We  also found that the one-jump velocity and density models have a great influence on the NH$_3$ line profiles.  \textnormal{To allow probing} of the hot core, observations of higher-energy ammonia transitions are needed for opacity reasons.
The fundamental 1$_0$--0$_0$ ortho transition 572~GHz shows a possible molecular outflow activity \textnormal{that probably is caused by bipolar outflows, but to rule out a champagne outflow is beyond the scope of this paper}. 

We derived ammonia abundances relative to molecular hydrogen in the inner, middle and outer region of the envelope to be $X_{\rm in}=1.8\times10^{-9}$, $X_{\rm mid}=1.2\times10^{-9}$ and $X_{\rm out}=3.0\times10^{-9}$, respectively. 
The overall low derived abundances suggest that the molecular gas is not yet in steady-state. 
Alternatively, \textnormal{our} assumption of equal gas and dust temperatures might break down in the low-density region. The enriched abundance in the outer region is most likely because of UV photo-desorption. In the inner region, the high temperature will partially release NH$_{3}$ from the water ice matrix on dust particles and increase the ammonia abundance. From modeling we derived an ammonia OPR of $\sim$~0.5, in agreement with recent findings in translucent interstellar gas \citep{Persson12}. 

Our derived \textnormal{accretion} rates are found to be $0.41-4.5\times10^{-2}$~[M$_{\odot}$~yr$^{-1}$], which is high enough to overcome the radiation pressure from G34. 
Even though \textnormal{the} observed transitions only probe the dynamics of the envelope and the outer parts of the hot core, the mass accretion profile shown in Fig.~\ref{fig:massaccretionprofile} may, however, indicate that the accretion already has halted in the inner part of the hot core. We speculate that \textnormal{this may be due to an outflow activity or an accretion disk, alternatively that we no longer can use ammonia to trace the mass close to the central source as it will be ionized/dissociated.}

Our results demonstrate that rotational transitions of ammonia seen in absorption toward  a strong far-infrared continuum source can successfully be used to probe infall velocities, physical and chemical properties in a variety of evolutionary stages of \textnormal{
MSF} cloud cores.

%
\begin{acknowledgements}
Thanks go to J. J\o rgensen and 
V. Taquet for comments in an early stage and to B. Mookerjea for sharing the 2 cm continuum image with us. MH, CMP, JHB, ESW and \AA H acknowledge generous support from the Swedish National Space Board. The research of AC was supported by the Lundbeck foundation. Research at Centre for Star and Planet Formation is funded by the Danish National Research Foundation.

The Herschel spacecraft was designed, built, tested, and launched under a contract to ESA managed by the Herschel/Planck Project team by an industrial consortium under the overall responsibility of the prime contractor Thales Alenia Space (Cannes), and including Astrium (Friedrichshafen) responsible for the payload module and for system testing at spacecraft level, Thales Alenia Space (Turin) responsible for the service module, and Astrium (Toulouse) responsible for the telescope, with in excess of a hundred subcontractors.

HIFI has been designed and built by a consortium of institutes and university departments from across Europe, Canada and the United States under the leadership of SRON Netherlands Institute for Space Research, Groningen, The Netherlands and with major contributions from Germany, France and the US. Consortium members are: Canada: CSA, U.Waterloo; France: CESR, LAB, LERMA, IRAM; Germany: KOSMA, MPIfR, MPS; Ireland, NUI Maynooth; Italy: ASI, IFSI-INAF, Osservatorio Astrofisico di Arcetri-INAF; Netherlands: SRON, TUD; Poland: CAMK, CBK; Spain: Observatorio Astron$\'o$mico Nacional (IGN), Centro de Astrobiolog$\'{\i}$a (CSIC-INTA). Sweden: Chalmers University of Technology - MC2, RSS \& GARD; Onsala Space Observatory; Swedish National Space Board, Stockholm University -Stockholm Observatory; Switzerland: ETH Zurich, FHNW; USA: Caltech, JPL, NHSC. 

This publication makes use of data products from the Wide-field Infrared Survey Explorer, which is a joint project of the University of California, Los Angeles, and the Jet Propulsion Laboratory/California Institute of Technology, funded by the National Aeronautics and Space Administration.

\end{acknowledgements}

\bibliographystyle{aa}
\bibliography{references}

\appendix

\section{Online Material: Details of the observations and examples of less successful models.}
\label{Appendix:obsID}
\onltab{
\begin{table*}[\!htb] 
\centering
\caption{G34.26+0.15: $Herschel$-HIFI OBSID's of the observed transitions analysed in this paper.
}
\begin{tabular} {lccccc} 
 \hline\hline
     \noalign{\smallskip}
Specie 	& Frequency  &  Band\tablefootmark{$a$} &	 LO-setting\tablefootmark{$b$}    & Date  &  OBSID  
\\    \noalign{\smallskip}	
&	 (GHz) & 	&	&     & 
\\     \noalign{\smallskip}
     \hline
\noalign{\smallskip}  
o-\element[][][][3]{NH}	& 	572.498  	&   1b (USB)   	&	A 	&  2011-04-22	& 1342219281 \\ \noalign{\smallskip}
& 		 		&   	   	&	B 	&  			& 1342219282 \\ \noalign{\smallskip}
& 		  		&   	   	&	C 	&  			& 1342219283 \\ \\
&	1214.852		&   5a (USB)   	&	A 	&  2011-03-13	& 1342215886 \\ \noalign{\smallskip}
& 		 		&   	   	&	B 	&  			& 1342215884 \\ \noalign{\smallskip}
& 		  		&   	   	&	C 	&  			& 1342215885 \\ \\
&	1763.524		&   7a (USB)  	&	A 	&  2012-04-20	& 1342244604  \\ \noalign{\smallskip}
& 		 		&   	   	&	B 	&  			& 1342244605 \\ \noalign{\smallskip}
& 		  		&   	   	&	C 	&  			& 1342244606 \\ \\
p-\element[][][][3]{NH}	& 	1215.245 		&    5a (USB)  	&	A 	&  2011-04-22	& 1342215886 \\ \noalign{\smallskip}
& 		 		&   	   	&	B 	&  			& 1342215884 \\ \noalign{\smallskip}
& 		  		&   	   	&	C 	&  			& 1342215885 \\ \\
&	1168.454		&    5a (LSB) 	&	A 	&  2012-04-18	& 1342244518 \\ \noalign{\smallskip}
& 		 		&   	   	&	B 	&  			& 1342244517 \\ \noalign{\smallskip}
& 		  		&   	   	&	C 	&  			& 1342244519 \\ \\
&	1763.824		&    7a (USB)  	&	A 	&  2011-04-20	& 1342244604  \\ \noalign{\smallskip}
& 		 		&   	   	&	B 	&  			& 1342244605 \\ \noalign{\smallskip}
& 		  		&   	   	&	C 	&  			& 1342244606 \\ \\
&	1763.602		&    7a (USB)  	&	A 	&  2011-04-20	& 1342244604  \\ \noalign{\smallskip}
& 		 		&   	   	&	B 	&  			& 1342244605 \\ \noalign{\smallskip}
& 		  		&   	   	&	C 	&  			& 1342244606 \\
\noalign{\smallskip} \noalign{\smallskip}
\hline 
\label{Table:obsid}
\end{tabular}
\tablefoot{
\tablefoottext{$a$}{HIFI consists of 7 different mixer bands and two double sideband
spectrometers (USB is the upper sideband, and LSB is the lower sideband).}
\tablefoottext{$b$}{Three different frequency settings of the LO were performed, with approximately 15 km$\,$s$^{-1}$ between each setting in order to determine the sideband origin of the signals.}
}
\end{table*} 
}

\onlfig{
\begin{figure}
\resizebox{\hsize}{!}{ 
\includegraphics{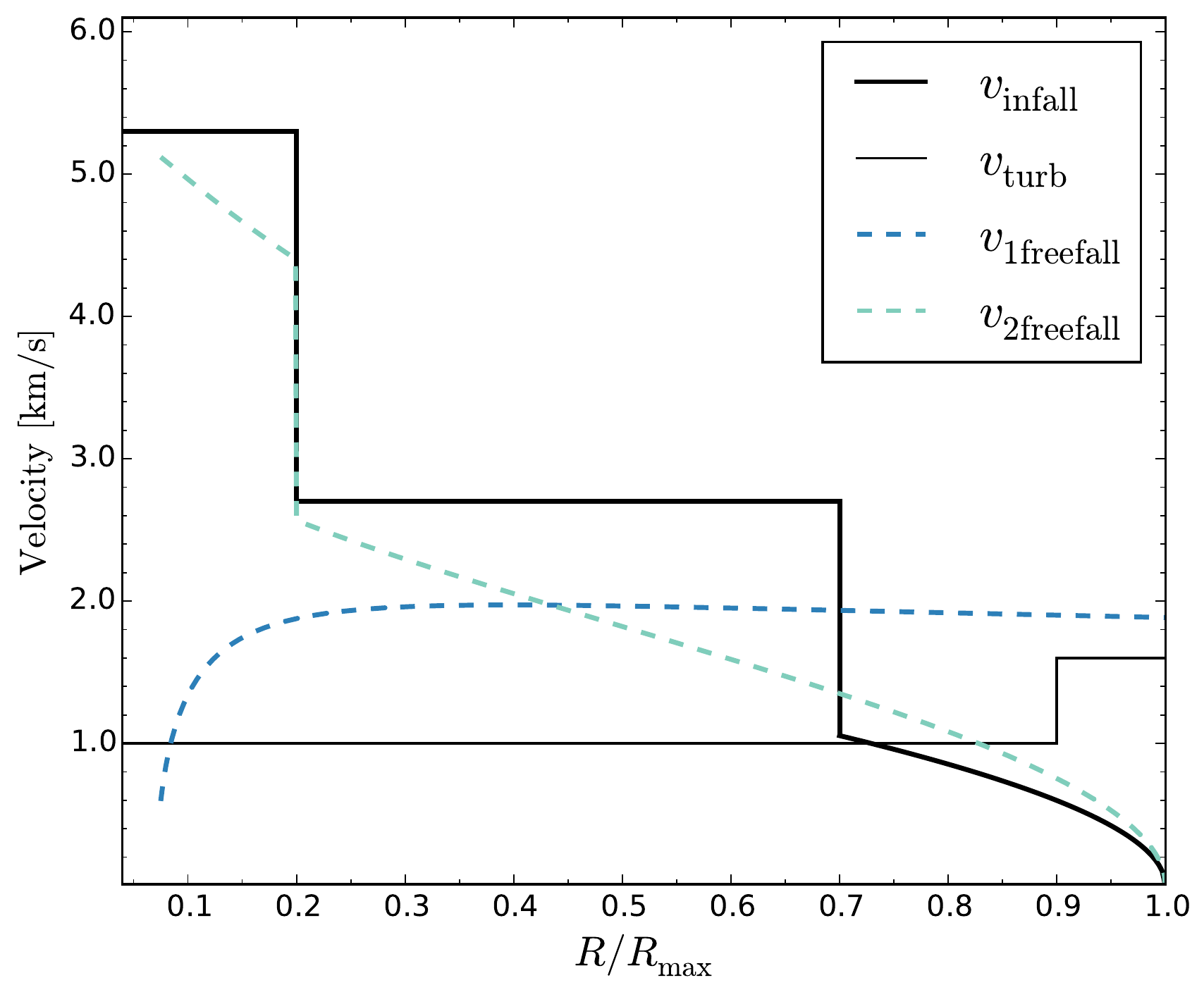}} 
\caption{Resulting velocity, $\varv_{\rm infall}$, and turbulent profile, $\varv_{\rm turb}$, as a function of the normalized radius of the cloud, obtained for the best-fitting ALI model (black solid lines). 
Velocity profile obtained for free-fall when modeling with one component (see Fig.~\ref{fig:fff0.3M20}),  $\varv_{\rm 1freefall}$  (blue dashed line), and two components (see Fig.~\ref{fig:fff0.6fff03M400}),  $\varv_{\rm 2freefall}$ (green dashed line). }
\label{fig:velprofile}
\end{figure} 
}

%

\onlfig{
 \begin{figure}[\!htb] 
\resizebox{\hsize}{!}{ 
\includegraphics{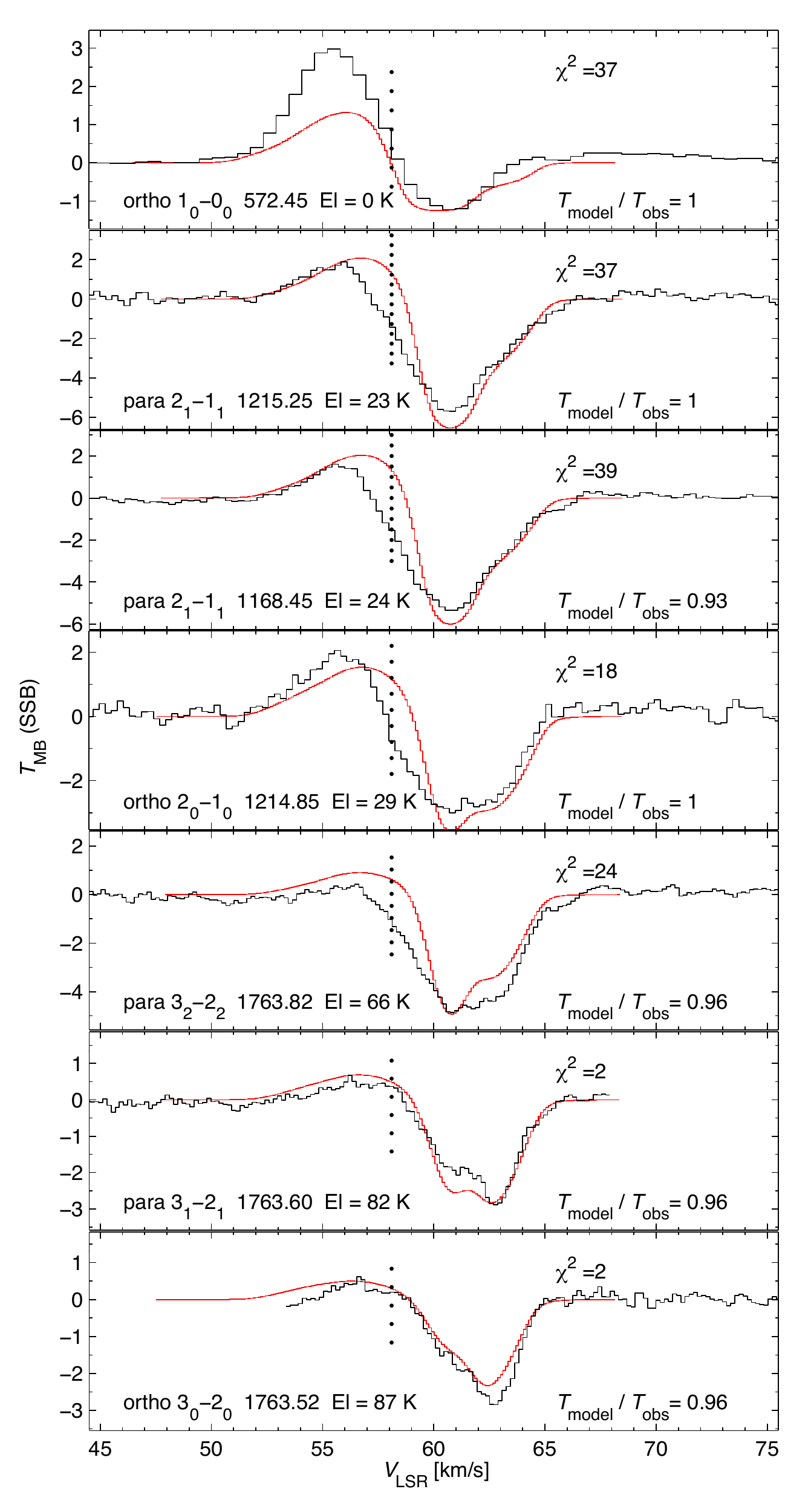}
}
\caption{ALI: 
\emph{Two velocity regimes with constant profiles} using 5.3~km$\,$s$^{-1}$ in the inner and 2.7~km$\,$s$^{-1}$ in the outer region. 
$\sum{\chi^{2}_{\rm line}}$~=~121 , over $\Delta V_{\mathrm{LSR}}= 54-68$~km$\,$s$^{-1}$. Notation as in Fig.~\ref{Fig: model}.
}
\label{Fig: 2constvel}
\end{figure} 
}

\onlfig{
 \begin{figure}
\resizebox{\hsize}{!}{ 
\includegraphics{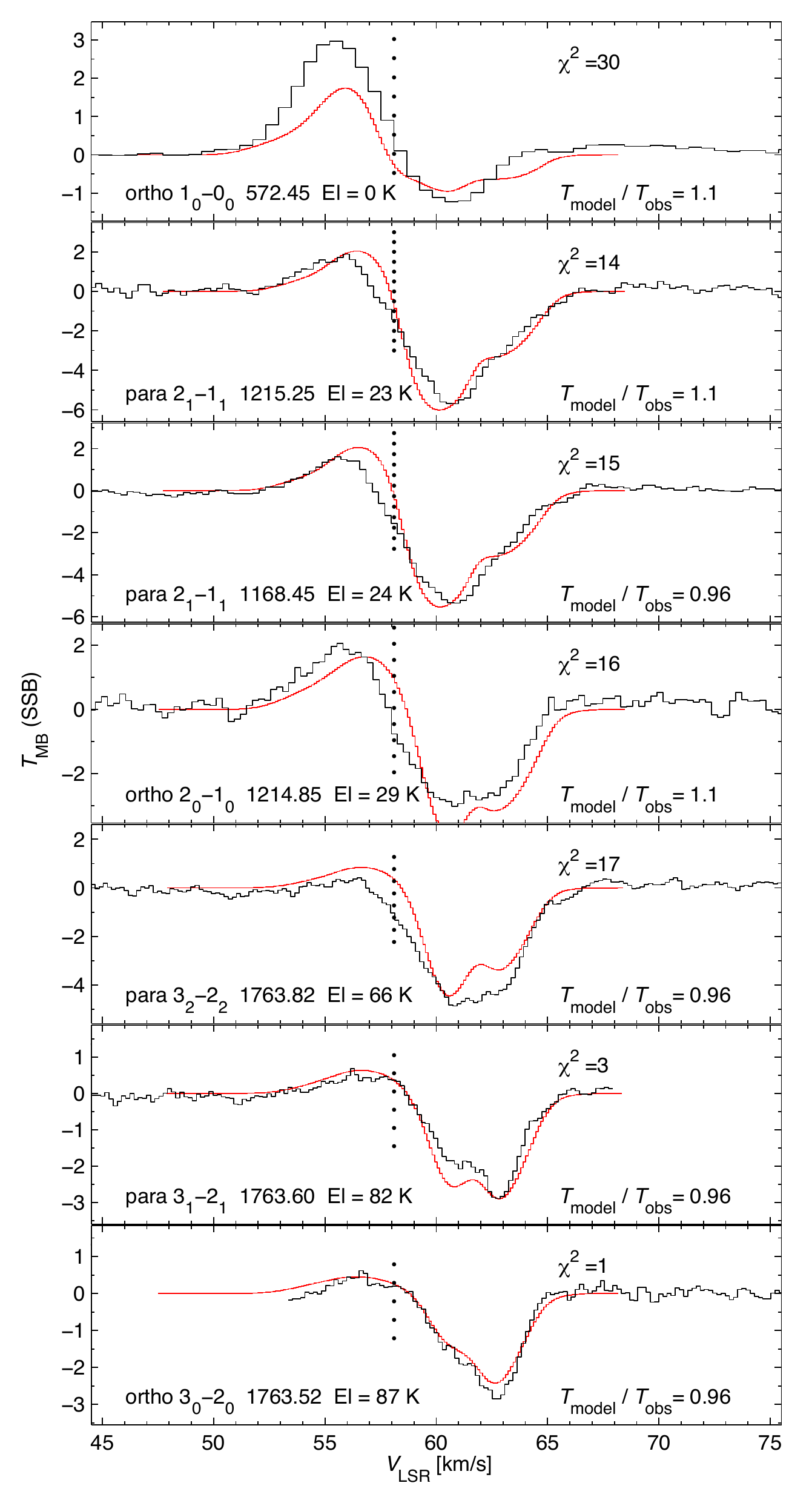}}
\caption{ALI: 
\emph{Two velocity regimes with free-fall} using $\mbox{$f$ = 0.7}$ in the inner and $\mbox{$f$ = 0.4}$ in the outer region, 
$\mbox{$M_0 = 400$~M$_{\odot}$}$ and $\mbox{$a$ = 1}$. 
Notation as in Fig. \ref{Fig: model}. $\mbox{$\sum{\chi^{2}_{\rm line}}$~=~65}$. 
}
\label{fig:fff0.6fff03M400}
\end{figure} 
}
\onlfig{
\begin{figure}
\resizebox{\hsize}{!}{ 
\includegraphics{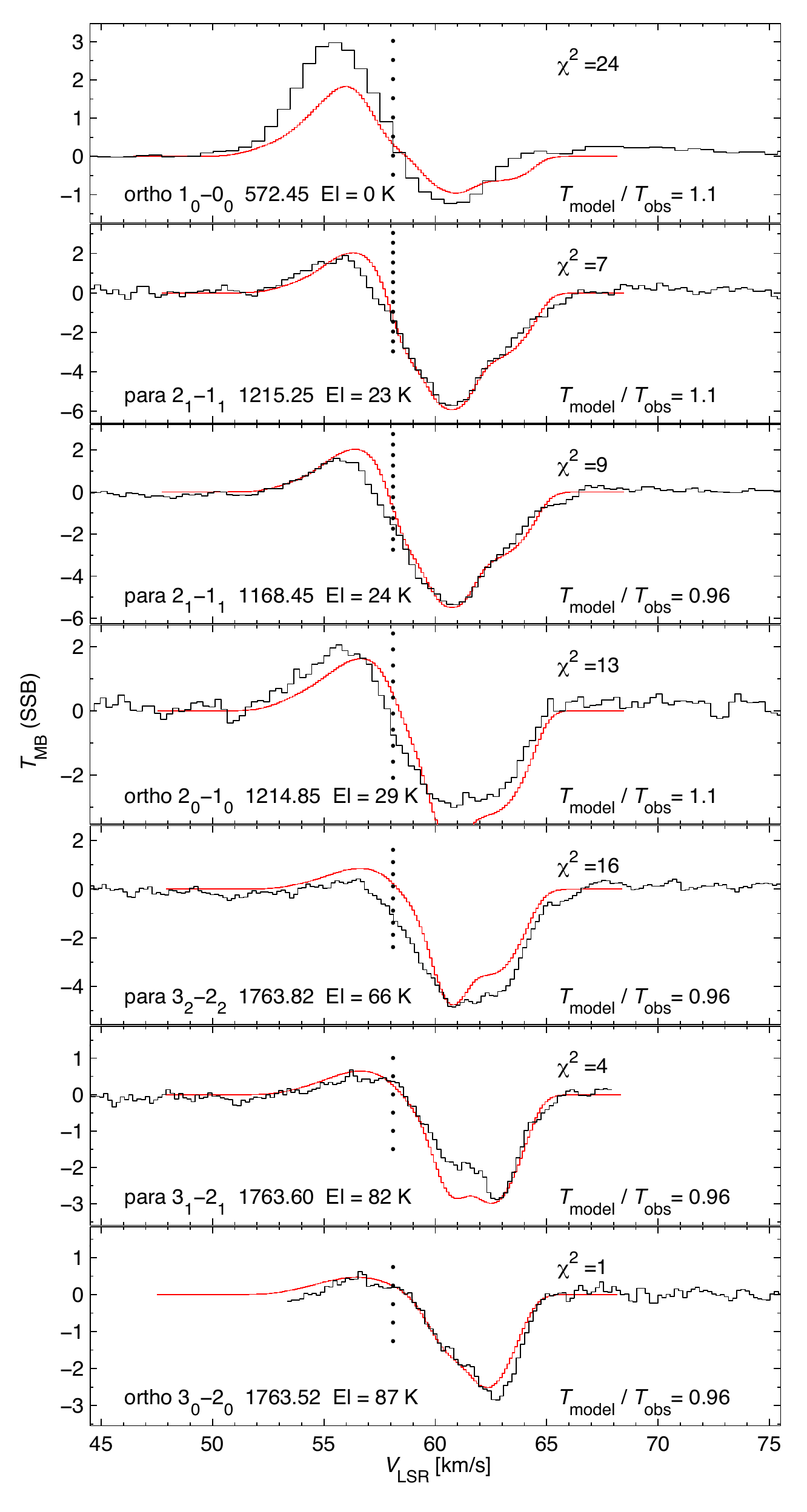}}
\caption{ALI: 
\emph{Three velocity regimes as in the best-fit model (Fig.~\ref{Fig: model}) with a radial continuous density profile. Notation as in } Fig.~\ref{Fig: model}. $\mbox{$\sum{\chi^{2}_{\rm line}}$ = 50}$.} 
\label{fig:constdens}
\end{figure} 
}

\onlfig{
\begin{figure}
\resizebox{\hsize}{!}{ 
\includegraphics{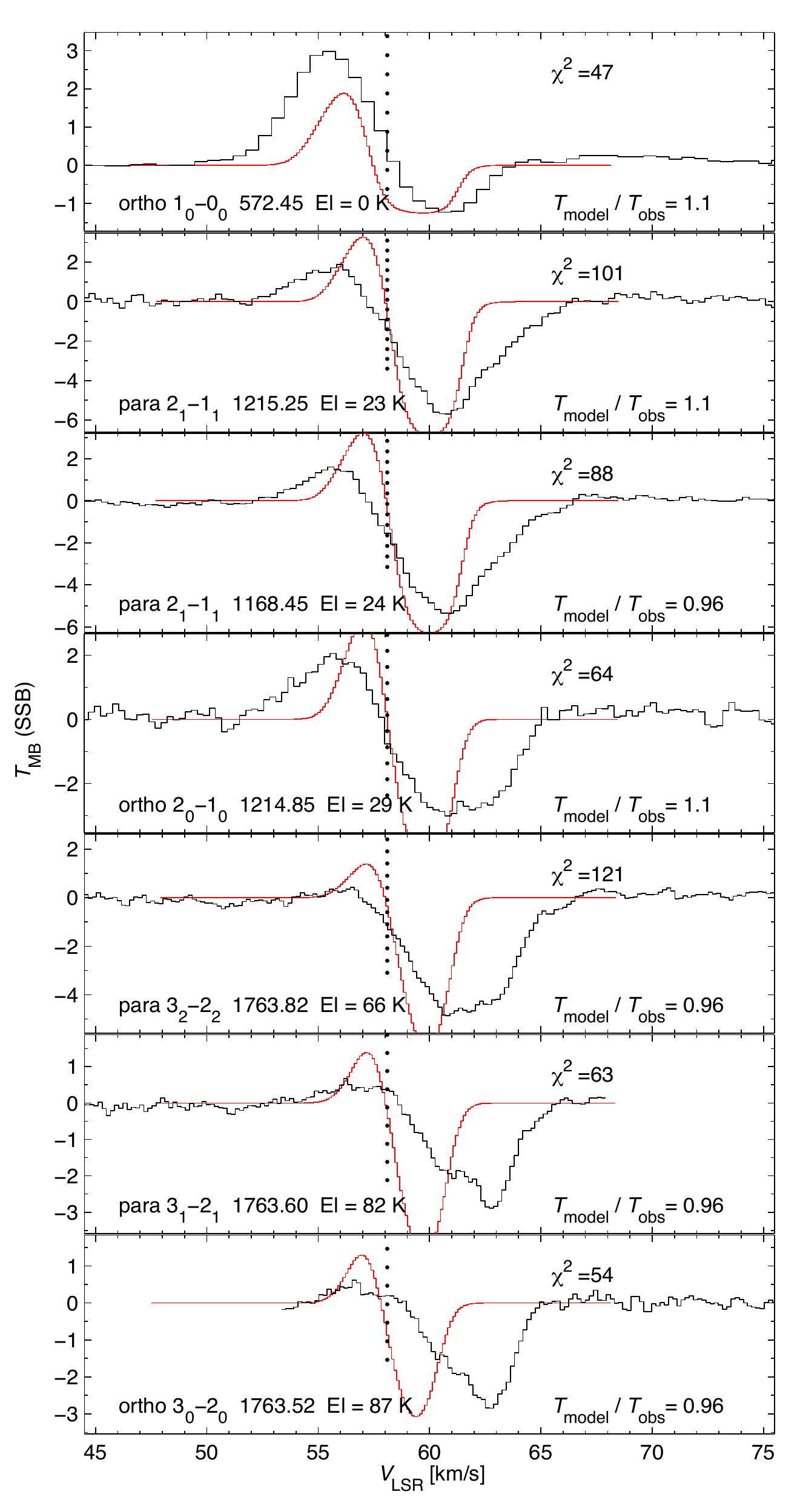}}
\caption{ALI: 
\emph{One velocity regime with free-fall} using $\mbox{$f$ = 0.3}$, $\mbox{$M_0 = 20$~M$_{\odot}$}$ and $\mbox{$a$ = 1000}$.
Notation as in Fig. \ref{Fig: model}. $\mbox{$\sum{\chi^{2}_{\rm line}}$ = 491}$. 
}
\label{fig:fff0.3M20}
\end{figure} 
}
\onlfig{
\begin{figure}
\resizebox{\hsize}{!}{ 
\includegraphics{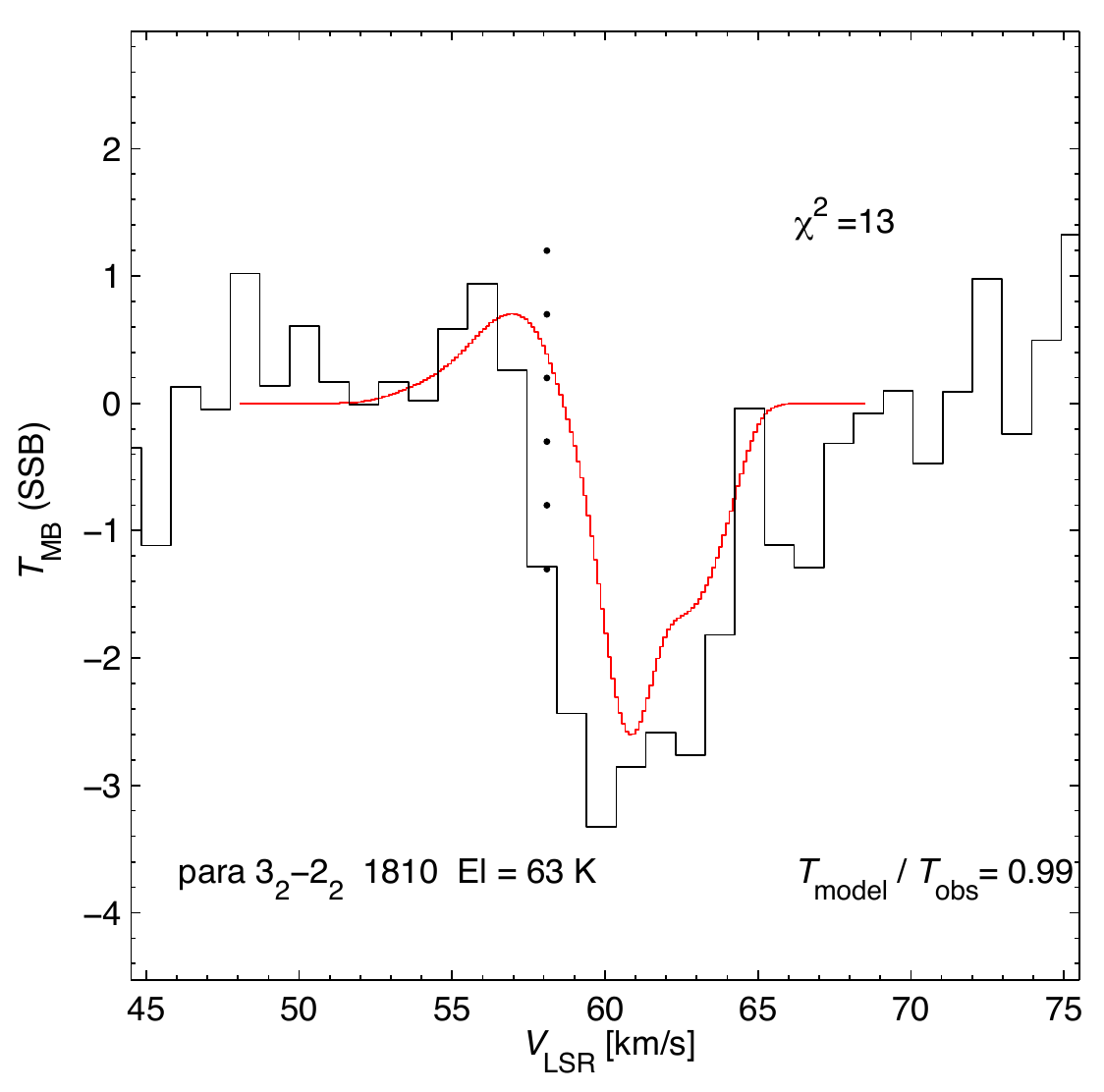}} 
\caption{ALI: 
Best-fit model applied on the 1810~GHz transition and multiplied by a factor of 30\% to reproduce the continuum. Notation as in  Fig.~\ref{Fig: model}.}
\label{Fig:bestfit1810}
\end{figure} 
}

\end{document}